\def\DIRvalue{RastelliRazamat}
\def\IDvalue{RR}
\def\titlevalue{The supersymmetric index in four dimensions}
\def\authorvalue{Leonardo Rastelli$^{1}$ and Shlomo S. Razamat$^{2}$}
\def\shortauthorvalue{Leonardo Rastelli and Shlomo S. Razamat}
\def\addressvalue{$^{1}$ C.~N.~Yang Institute for Theoretical Physics, Stony
  Brook University, Stony Brook, NY 11794-3840, USA\\
$^{2}$ Department of Physics, Technion, Haifa, 32000, Israel\\
  \tt leonardo.rastelli@stonybrook.edu, razamat@physics.technion.ac.il}

\def\abstractvalue{We review  the calculation and properties of the supersymmetric index for four dimensional ${\cal N}=1$ theories,
illustrating its physical significance  in several examples.}

\def\preprintvalue{}

\ifx\ifLONG\undefined \def\workingdir{.}

\documentclass[12pt]{article}
\newcommand{\chapterauthor}[1]{
\begin{center}
{\bf \normalsize  #1}
\end{center}
}

\newcommand{\chapteraddress}[1]{
\begin{center}
{ \small \it \addressvalue}
\end{center}
}

\newcommand{\chapterabstract}[1]{
\vspace{\baselineskip}
\begin{center}
\textbf{\small Abstract}
\end{center}
#1}

\newcommand{\chapterheader}{

\chapter[\titlevalue{}  (by \shortauthorvalue)]{\titlevalue}
\label{Chapter\IDvalue}
\chapterauthor{\authorvalue}
\chapteraddress{\addressvalue}
\chapterabstract{\abstractvalue}
\tightmtctrue
\minitoc
}

\newcommand{\documentheader}{
\begin{flushright} \small
  \preprintvalue
 \end{flushright}

\begin{center}
{\bf \Large \titlevalue}
\end{center}

\chapterauthor{\authorvalue}
\chapteraddress{\addressvalue}
\chapterabstract{\abstractvalue}

\medskip

This is a contribution to the review volume ``Localization techniques
in quantum field theories'' (eds. V.~Pestun and M.~Zabzine) which
contains 17 Chapters available at \cite{ContributionSummary}

\tableofcontents
}

\newcommand{\ifvolume}[2]{\ifx\ifLONG\undefined#2\else#1\fi}

\newcommand{\documentfinish}{
\ifx\ifLONG\undefined
\bibliographystyle{bibreview} 
\bibliography{\IDvalue,review}  
\end{document}
\else
\addcontentsline{toc}{section}{References}
\providecommand{\href}[2]{#2}\begingroup\raggedright\begin{thebibliography}{10}

\bibitem{ContributionSummary}
V.~Pestun and M.~Zabzine, eds., {\em Localization techniques in quantum field
  theory}, vol.~xx.
\newblock Journal of Physics A, 2016.
\newblock \href{http://arxiv.org/abs/1608.02952}{{\tt 1608.02952}}.
\newblock \url{https://arxiv.org/src/1608.02952/anc/LocQFT.pdf},
  \url{http://pestun.ihes.fr/pages/LocalizationReview/LocQFT.pdf}.

\bibitem{RRPestun:2007rz}
V.~Pestun, ``{Localization of gauge theory on a four-sphere and supersymmetric
  Wilson loops},''
\href{http://arxiv.org/abs/0712.2824}{{\tt arXiv:0712.2824 [hep-th]}}.

\bibitem{ContributionHO}
K.~Hosomichi, ``$\mathcal{N}=2$ SUSY gauge theories on $S^4$,'' {\em Journal of
  Physics A} {\bf xx} (2016)  000, \href{http://arxiv.org/abs/1608.02962}{{\tt
  1608.02962}}.

\bibitem{RRKinney:2005ej}
J.~Kinney, J.~M. Maldacena, S.~Minwalla, and S.~Raju, ``{An index for 4
  dimensional super conformal theories},''
  \href{http://dx.doi.org/10.1007/s00220-007-0258-7}{{\em Commun. Math. Phys.}
  {\bf 275} (2007)  209--254},
\href{http://arxiv.org/abs/hep-th/0510251}{{\tt arXiv:hep-th/0510251}}.

\bibitem{RRRomelsberger:2005eg}
C.~Romelsberger, ``{Counting chiral primaries in N = 1, d=4 superconformal
  field theories},''
  \href{http://dx.doi.org/10.1016/j.nuclphysb.2006.03.037}{{\em Nucl. Phys.}
  {\bf B747} (2006)  329--353},
\href{http://arxiv.org/abs/hep-th/0510060}{{\tt arXiv:hep-th/0510060}}.

\bibitem{RRRomelsberger:2007ec}
C.~Romelsberger, ``{Calculating the Superconformal Index and Seiberg
  Duality},''
\href{http://arxiv.org/abs/0707.3702}{{\tt arXiv:0707.3702 [hep-th]}}.

\bibitem{RRZwiebel:2011wa}
B.~I. Zwiebel, ``{Charging the Superconformal Index},''
  \href{http://dx.doi.org/10.1007/JHEP01(2012)116}{{\em JHEP} {\bf 01} (2012)
  116},
\href{http://arxiv.org/abs/1111.1773}{{\tt arXiv:1111.1773 [hep-th]}}.

\bibitem{RRFestuccia:2011ws}
G.~Festuccia and N.~Seiberg, ``{Rigid Supersymmetric Theories in Curved
  Superspace},'' \href{http://dx.doi.org/10.1007/JHEP06(2011)114}{{\em JHEP}
  {\bf 1106} (2011)  114}, \href{http://arxiv.org/abs/1105.0689}{{\tt
  arXiv:1105.0689 [hep-th]}}.

\bibitem{RRSen:1985ph}
D.~Sen, ``{Supersymmetry in the Space-time R X S**3},''
\href{http://dx.doi.org/10.1016/0550-3213(87)90033-2}{{\em Nucl. Phys.} {\bf
  B284} (1987)  201}.

\bibitem{ContributionDU}
T.~Dumitrescu, ``An Introduction to Supersymmetric Field Theories in Curved
  Space,'' {\em Journal of Physics A} {\bf xx} (2016)  000,
  \href{http://arxiv.org/abs/1608.02957}{{\tt 1608.02957}}.

\bibitem{RRClosset:2013sxa}
C.~Closset and I.~Shamir, ``{The $\mathcal{N}=1$ Chiral Multiplet on $T^2\times
  S^2$ and Supersymmetric Localization},''
  \href{http://dx.doi.org/10.1007/JHEP03(2014)040}{{\em JHEP} {\bf 03} (2014)
  040},
\href{http://arxiv.org/abs/1311.2430}{{\tt arXiv:1311.2430 [hep-th]}}.

\bibitem{RRAssel:2014paa}
B.~Assel, D.~Cassani, and D.~Martelli, ``{Localization on Hopf surfaces},''
  \href{http://dx.doi.org/10.1007/JHEP08(2014)123}{{\em JHEP} {\bf 08} (2014)
  123},
\href{http://arxiv.org/abs/1405.5144}{{\tt arXiv:1405.5144 [hep-th]}}.

\bibitem{RRKim:2012ava}
H.-C. Kim and S.~Kim, ``{M5-branes from gauge theories on the 5-sphere},''
\href{http://arxiv.org/abs/1206.6339}{{\tt arXiv:1206.6339 [hep-th]}}.

\bibitem{RRArdehali:2015bla}
A.~A. Ardehali, ``{High-temperature asymptotics of supersymmetric partition
  functions},''
\href{http://arxiv.org/abs/1512.03376}{{\tt arXiv:1512.03376 [hep-th]}}.

\bibitem{RRAssel:2015nca}
B.~Assel, D.~Cassani, L.~Di~Pietro, Z.~Komargodski, J.~Lorenzen, and
  D.~Martelli, ``{The Casimir Energy in Curved Space and its Supersymmetric
  Counterpart},'' \href{http://dx.doi.org/10.1007/JHEP07(2015)043}{{\em JHEP}
  {\bf 07} (2015)  043},
\href{http://arxiv.org/abs/1503.05537}{{\tt arXiv:1503.05537 [hep-th]}}.

\bibitem{RRBobev:2015kza}
N.~Bobev, M.~Bullimore, and H.-C. Kim, ``{Supersymmetric Casimir Energy and the
  Anomaly Polynomial},'' \href{http://dx.doi.org/10.1007/JHEP09(2015)142}{{\em
  JHEP} {\bf 09} (2015)  142},
\href{http://arxiv.org/abs/1507.08553}{{\tt arXiv:1507.08553 [hep-th]}}.

\bibitem{RRDolan:2008qi}
F.~A. Dolan and H.~Osborn, ``{Applications of the Superconformal Index for
  Protected Operators and q-Hypergeometric Identities to N=1 Dual Theories},''
  \href{http://dx.doi.org/10.1016/j.nuclphysb.2009.01.028}{{\em Nucl. Phys.}
  {\bf B818} (2009)  137--178},
\href{http://arxiv.org/abs/arXiv:0801.4947}{{\tt arXiv:arXiv:0801.4947
  [hep-th]}}.

\bibitem{RRBenvenuti:2006qr}
S.~Benvenuti, B.~Feng, A.~Hanany, and Y.-H. He, ``{Counting BPS operators in
  gauge theories: Quivers, syzygies and plethystics},''
  \href{http://dx.doi.org/10.1088/1126-6708/2007/11/050}{{\em JHEP} {\bf 11}
  (2007)  050},
\href{http://arxiv.org/abs/hep-th/0608050}{{\tt arXiv:hep-th/0608050}}.

\bibitem{RRAharony:2003sx}
O.~Aharony, J.~Marsano, S.~Minwalla, K.~Papadodimas, and M.~Van~Raamsdonk,
  ``{The Hagedorn / deconfinement phase transition in weakly coupled large N
  gauge theories},'' {\em Adv. Theor. Math. Phys.} {\bf 8} (2004)  603--696,
\href{http://arxiv.org/abs/hep-th/0310285}{{\tt arXiv:hep-th/0310285}}.

\bibitem{RRGadde:2009kb}
A.~Gadde, E.~Pomoni, L.~Rastelli, and S.~S. Razamat, ``{S-duality and 2d
  Topological QFT},'' \href{http://dx.doi.org/10.1007/JHEP03(2010)032}{{\em
  JHEP} {\bf 03} (2010)  032},
\href{http://arxiv.org/abs/0910.2225}{{\tt arXiv:0910.2225 [hep-th]}}.

\bibitem{RRAharony:2013dha}
O.~Aharony, S.~S. Razamat, N.~Seiberg, and B.~Willett, ``{3d dualities from 4d
  dualities},'' \href{http://dx.doi.org/10.1007/JHEP07(2013)149}{{\em JHEP}
  {\bf 07} (2013)  149},
\href{http://arxiv.org/abs/1305.3924}{{\tt arXiv:1305.3924 [hep-th]}}.

\bibitem{RRSeiberg:1994pq}
N.~Seiberg, ``{Electric - magnetic duality in supersymmetric nonAbelian gauge
  theories},'' \href{http://dx.doi.org/10.1016/0550-3213(94)00023-8}{{\em Nucl.
  Phys.} {\bf B435} (1995)  129--146},
\href{http://arxiv.org/abs/hep-th/9411149}{{\tt arXiv:hep-th/9411149
  [hep-th]}}.

\bibitem{spibrb}
V.~Spiridonov, ``{On the elliptic beta function},'' {\em Russian Mathematical
  Surveys} {\bf 56:1} (2001)  185.

\bibitem{RRSpiridonov2}
V.~Spiridonov, ``{Essays on the theory of elliptic hypergeometric functions},''
  {\em Uspekhi Mat. Nauk} {\bf 63 no 3} (2008)  3--72,
  \href{http://arxiv.org/abs/arXiv:0805.3135}{{\tt arXiv:0805.3135}}.

\bibitem{RRrains}
E.~M. Rains, ``{Transformations of Elliptic Hypergeometric Integrals },''
  \href{http://arxiv.org/abs/math/0309252}{{\tt math/0309252}}.

\bibitem{RRSpiridonov:2012ww}
V.~Spiridonov and G.~Vartanov, ``{Elliptic hypergeometric integrals and 't
  Hooft anomaly matching conditions},''
  \href{http://dx.doi.org/10.1007/JHEP06(2012)016}{{\em JHEP} {\bf 1206} (2012)
   016},
\href{http://arxiv.org/abs/1203.5677}{{\tt arXiv:1203.5677 [hep-th]}}.

\bibitem{RRspirinv}
V.~P. Spiridonov and S.~O. Warnaar, ``{Inversions of integral operators and
  elliptic beta integrals on root systems },'' {\em Adv. Math.} {\bf 207}
  (2006)  91--132, \href{http://arxiv.org/abs/math/0411044}{{\tt
  math/0411044}}.

\bibitem{RRBeem:2012yn}
C.~Beem and A.~Gadde, ``{The $N=1$ superconformal index for class $S$ fixed
  points},'' \href{http://dx.doi.org/10.1007/JHEP04(2014)036}{{\em JHEP} {\bf
  1404} (2014)  036},
\href{http://arxiv.org/abs/1212.1467}{{\tt arXiv:1212.1467 [hep-th]}}.

\bibitem{RRGreen:2010da}
D.~Green, Z.~Komargodski, N.~Seiberg, Y.~Tachikawa, and B.~Wecht, ``{Exactly
  Marginal Deformations and Global Symmetries},''
  \href{http://dx.doi.org/10.1007/JHEP06(2010)106}{{\em JHEP} {\bf 06} (2010)
  106},
\href{http://arxiv.org/abs/1005.3546}{{\tt arXiv:1005.3546 [hep-th]}}.

\bibitem{RRFocco}
F.~J. van~de Bult, ``An elliptic hypergeometric integral with $W(F_4)$
  symmetry,'' \href{http://arxiv.org/abs/arXiv:0909.4793}{{\tt
  arXiv:0909.4793}}.

\bibitem{RRSpiridonov:2008zr}
V.~P. Spiridonov and G.~S. Vartanov, ``{Superconformal indices for ${\mathcal
  N}=1$ theories with multiple duals},''
  \href{http://dx.doi.org/10.1016/j.nuclphysb.2009.08.022}{{\em Nucl. Phys.}
  {\bf B824} (2010)  192--216},
\href{http://arxiv.org/abs/0811.1909}{{\tt arXiv:0811.1909 [hep-th]}}.

\bibitem{RRDimofte:2012pd}
T.~Dimofte and D.~Gaiotto, ``{An E7 Surprise},''
  \href{http://dx.doi.org/10.1007/JHEP10(2012)129}{{\em JHEP} {\bf 10} (2012)
  129},
\href{http://arxiv.org/abs/1209.1404}{{\tt arXiv:1209.1404 [hep-th]}}.

\bibitem{RRGaiotto:2012xa}
D.~Gaiotto, L.~Rastelli, and S.~S. Razamat, ``{Bootstrapping the superconformal
  index with surface defects},'' {\em JHEP} {\bf 01} (2013)  022,
\href{http://arxiv.org/abs/1207.3577}{{\tt arXiv:1207.3577 [hep-th]}}.

\bibitem{RR2014arXiv1412.2781G}
D.~{Gaiotto} and H.-C. {Kim}, ``{Surface defects and instanton partition
  functions},'' {\em ArXiv e-prints} (2014)  ,
  \href{http://arxiv.org/abs/1412.2781}{{\tt arXiv:1412.2781 [hep-th]}}.

\bibitem{RR2014JHEP...03..080G}
A.~{Gadde} and S.~{Gukov}, ``{2d index and surface operators},''
  \href{http://dx.doi.org/10.1007/JHEP03(2014)080}{{\em Journal of High Energy
  Physics} {\bf 3} (2014)  80}, \href{http://arxiv.org/abs/1305.0266}{{\tt
  arXiv:1305.0266 [hep-th]}}.

\bibitem{RRSpiridonov:2010qv}
V.~Spiridonov and G.~Vartanov, ``{Superconformal indices of N=4 SYM field
  theories},'' \href{http://arxiv.org/abs/1005.4196}{{\tt arXiv:1005.4196
  [hep-th]}}.

\bibitem{RRKutasov:1995ve}
D.~Kutasov, ``{A Comment on duality in N=1 supersymmetric nonAbelian gauge
  theories},'' \href{http://dx.doi.org/10.1016/0370-2693(95)00392-X}{{\em Phys.
  Lett.} {\bf B351} (1995)  230--234},
\href{http://arxiv.org/abs/hep-th/9503086}{{\tt arXiv:hep-th/9503086
  [hep-th]}}.

\bibitem{RRKutasov:1995np}
D.~Kutasov and A.~Schwimmer, ``{On duality in supersymmetric Yang-Mills
  theory},'' \href{http://dx.doi.org/10.1016/0370-2693(95)00676-C}{{\em Phys.
  Lett.} {\bf B354} (1995)  315--321},
\href{http://arxiv.org/abs/hep-th/9505004}{{\tt arXiv:hep-th/9505004
  [hep-th]}}.

\bibitem{RRIntriligator:2003mi}
K.~A. Intriligator and B.~Wecht, ``{RG fixed points and flows in SQCD with
  adjoints},'' \href{http://dx.doi.org/10.1016/j.nuclphysb.2003.10.033}{{\em
  Nucl. Phys.} {\bf B677} (2004)  223--272},
\href{http://arxiv.org/abs/hep-th/0309201}{{\tt arXiv:hep-th/0309201
  [hep-th]}}.

\bibitem{RRKutasov:2014yqa}
D.~Kutasov and J.~Lin, ``{Exceptional N=1 Duality},''
\href{http://arxiv.org/abs/1401.4168}{{\tt arXiv:1401.4168 [hep-th]}}.

\bibitem{RRKutasov:2014wwa}
D.~Kutasov and J.~Lin, ``{N=1 Duality and the Superconformal Index},''
\href{http://arxiv.org/abs/1402.5411}{{\tt arXiv:1402.5411 [hep-th]}}.

\bibitem{RRSpiridonov:2009za}
V.~P. Spiridonov and G.~S. Vartanov, ``{Elliptic hypergeometry of
  supersymmetric dualities},''
\href{http://arxiv.org/abs/0910.5944}{{\tt arXiv:0910.5944 [hep-th]}}.

\bibitem{RRSpiridonov:2011hf}
V.~Spiridonov and G.~Vartanov, ``{Elliptic hypergeometry of supersymmetric
  dualities II. Orthogonal groups, knots, and vortices},''
  \href{http://arxiv.org/abs/1107.5788}{{\tt arXiv:1107.5788 [hep-th]}}.

\bibitem{RRSpiridonov:2010yc}
V.~Spiridonov, ``{Elliptic hypergeometric terms},''
  \href{http://arxiv.org/abs/1003.4491}{{\tt arXiv:1003.4491 [math.CA]}}.

\bibitem{RRSpiridonov3}
V.~Spiridonov, ``{Classical elliptic hypergeometric functions and their
  applications },'' {\em Rokko Lect. in Math. Vol. 18, Dept. of Math, Kobe
  Univ.} (2005)  253--287, \href{http://arxiv.org/abs/arXiv:math/0511579}{{\tt
  arXiv:math/0511579}}.

\bibitem{RRSpiridonov4}
V.~Spiridonov, ``{Elliptic hypergeometric functions},''
  \href{http://arxiv.org/abs/arXiv:0704.3099}{{\tt arXiv:0704.3099}}.

\bibitem{RRClosset:2012ru}
C.~Closset, T.~T. Dumitrescu, G.~Festuccia, and Z.~Komargodski,
  ``{Supersymmetric Field Theories on Three-Manifolds},''
  \href{http://dx.doi.org/10.1007/JHEP05(2013)017}{{\em JHEP} {\bf 05} (2013)
  017},
\href{http://arxiv.org/abs/1212.3388}{{\tt arXiv:1212.3388 [hep-th]}}.

\bibitem{RRAharony:2013kma}
O.~Aharony, S.~S. Razamat, N.~Seiberg, and B.~Willett, ``{3$d$ dualities from
  4$d$ dualities for orthogonal groups},''
  \href{http://dx.doi.org/10.1007/JHEP08(2013)099}{{\em JHEP} {\bf 1308} (2013)
   099},
\href{http://arxiv.org/abs/1307.0511}{{\tt arXiv:1307.0511}}.

\bibitem{RRIntriligator:1994rx}
K.~A. Intriligator, N.~Seiberg, and S.~H. Shenker, ``{Proposal for a simple
  model of dynamical SUSY breaking},''
  \href{http://dx.doi.org/10.1016/0370-2693(94)01336-B}{{\em Phys. Lett.} {\bf
  B342} (1995)  152--154},
\href{http://arxiv.org/abs/hep-ph/9410203}{{\tt arXiv:hep-ph/9410203
  [hep-ph]}}.

\bibitem{RRVartanov:2010xj}
G.~Vartanov, ``{On the ISS model of dynamical SUSY breaking},''
  \href{http://dx.doi.org/10.1016/j.physletb.2010.12.040}{{\em Phys.Lett.} {\bf
  B696} (2011)  288--290}, \href{http://arxiv.org/abs/1009.2153}{{\tt
  arXiv:1009.2153 [hep-th]}}.

\bibitem{RRImamura:2011wg}
Y.~Imamura and D.~Yokoyama, ``{N=2 supersymmetric theories on squashed
  three-sphere},'' \href{http://dx.doi.org/10.1103/PhysRevD.85.025015}{{\em
  Phys. Rev.} {\bf D85} (2012)  025015},
\href{http://arxiv.org/abs/1109.4734}{{\tt arXiv:1109.4734 [hep-th]}}.

\bibitem{RRfelvar}
G.~{Felder} and A.~{Varchenko}, ``{The elliptic gamma function and $SL(3,Z)
  \times {\mathbb Z}^3$},'' {\em ArXiv Mathematics e-prints} (1999)  ,
  \href{http://arxiv.org/abs/math/9907061}{{\tt math/9907061}}.

\bibitem{RRKapustin:2009kz}
A.~Kapustin, B.~Willett, and I.~Yaakov, ``{Exact Results for Wilson Loops in
  Superconformal Chern- Simons Theories with Matter},''
  \href{http://dx.doi.org/10.1007/JHEP03(2010)089}{{\em JHEP} {\bf 03} (2010)
  089},
\href{http://arxiv.org/abs/0909.4559}{{\tt arXiv:0909.4559 [hep-th]}}.

\bibitem{RRDolan:2011rp}
F.~Dolan, V.~Spiridonov, and G.~Vartanov, ``{From 4d superconformal indices to
  3d partition functions},'' \href{http://arxiv.org/abs/1104.1787}{{\tt
  arXiv:1104.1787 [hep-th]}}.

\bibitem{RRImamura:2011uw}
Y.~Imamura, ``{Relation between the 4d superconformal index and the $S^3$
  partition function},'' \href{http://arxiv.org/abs/1104.4482}{{\tt
  arXiv:1104.4482 [hep-th]}}.

\bibitem{RRGadde:2011ia}
A.~Gadde and W.~Yan, ``{Reducing the 4d Index to the $S^3$ Partition
  Function},'' \href{http://dx.doi.org/10.1007/JHEP12(2012)003}{{\em JHEP} {\bf
  1212} (2012)  003},
\href{http://arxiv.org/abs/1104.2592}{{\tt arXiv:1104.2592 [hep-th]}}.

\bibitem{ContributionWI}
B.~Willett, ``Localization on three-dimensional manifolds,'' {\em Journal of
  Physics A} {\bf xx} (2016)  000, \href{http://arxiv.org/abs/1608.02958}{{\tt
  1608.02958}}.

\bibitem{RRGray:2008yu}
J.~Gray, A.~Hanany, Y.-H. He, V.~Jejjala, and N.~Mekareeya, ``{SQCD: A
  Geometric Apercu},''
  \href{http://dx.doi.org/10.1088/1126-6708/2008/05/099}{{\em JHEP} {\bf 0805}
  (2008)  099}, \href{http://arxiv.org/abs/0803.4257}{{\tt arXiv:0803.4257
  [hep-th]}}.

\bibitem{RRHanany:2008kn}
A.~Hanany and N.~Mekareeya, ``{Counting Gauge Invariant Operators in SQCD with
  Classical Gauge Groups},''
  \href{http://dx.doi.org/10.1088/1126-6708/2008/10/012}{{\em JHEP} {\bf 10}
  (2008)  012},
\href{http://arxiv.org/abs/0805.3728}{{\tt arXiv:0805.3728 [hep-th]}}.

\bibitem{RRGadde:2011uv}
A.~Gadde, L.~Rastelli, S.~S. Razamat, and W.~Yan, ``{Gauge Theories and
  Macdonald Polynomials},''
  \href{http://dx.doi.org/10.1007/s00220-012-1607-8}{{\em Commun.Math.Phys.}
  {\bf 319} (2013)  147--193},
\href{http://arxiv.org/abs/1110.3740}{{\tt arXiv:1110.3740 [hep-th]}}.

\bibitem{RRGaiotto:2009we}
D.~Gaiotto, ``{N=2 dualities},''
\href{http://arxiv.org/abs/arXiv:0904.2715}{{\tt arXiv:arXiv:0904.2715
  [hep-th]}}.

\bibitem{RRRazamat:2014pta}
S.~S. Razamat and B.~Willett, ``{Down the rabbit hole with theories of class
  S},''
\href{http://arxiv.org/abs/1403.6107}{{\tt arXiv:1403.6107 [hep-th]}}.

\bibitem{RRCremonesi:2015dja}
S.~Cremonesi, ``{The Hilbert series of 3d ${\boldsymbol{\mathcal{N}}}=2$
  Yang-Mills theories with vectorlike matter},''
  \href{http://dx.doi.org/10.1088/1751-8113/48/45/455401}{{\em J. Phys.} {\bf
  A48} (2015) no.~45, 455401},
\href{http://arxiv.org/abs/1505.02409}{{\tt arXiv:1505.02409 [hep-th]}}.

\bibitem{RRHanany:2015via}
A.~Hanany, C.~Hwang, H.~Kim, J.~Park, and R.-K. Seong, ``{Hilbert Series for
  Theories with Aharony Duals},''
  \href{http://dx.doi.org/10.1007/JHEP11(2015)132,
  10.1007/JHEP04(2016)064}{{\em JHEP} {\bf 11} (2015)  132},
  \href{http://arxiv.org/abs/1505.02160}{{\tt arXiv:1505.02160 [hep-th]}}.
[Addendum: JHEP04,064(2016)].

\bibitem{RRGaiotto:2015usa}
D.~Gaiotto and S.~S. Razamat, ``{$ \mathcal{N}=1 $ theories of class $
  {\mathcal{S}}_k $},'' \href{http://dx.doi.org/10.1007/JHEP07(2015)073}{{\em
  JHEP} {\bf 07} (2015)  073},
\href{http://arxiv.org/abs/1503.05159}{{\tt arXiv:1503.05159 [hep-th]}}.

\bibitem{RRRastelli:2014jja}
L.~Rastelli and S.~S. Razamat,
  \href{http://dx.doi.org/10.1007/978-3-319-18769-3_9}{``{The Superconformal
  Index of Theories of Class $\mathcal {S}$},''} in {\em New Dualities of
  Supersymmetric Gauge Theories}, J.~Teschner, ed., pp.~261--305.
\newblock 2016.
\newblock
\href{http://arxiv.org/abs/1412.7131}{{\tt arXiv:1412.7131 [hep-th]}}.
\newblock

\bibitem{RRAlday:2013kda}
L.~F. Alday, M.~Bullimore, M.~Fluder, and L.~Hollands, ``{Surface defects, the
  superconformal index and q-deformed Yang-Mills},''
\href{http://arxiv.org/abs/1303.4460}{{\tt arXiv:1303.4460 [hep-th]}}.

\bibitem{RRGadde:2010en}
A.~Gadde, L.~Rastelli, S.~S. Razamat, and W.~Yan, ``{On the Superconformal
  Index of N=1 IR Fixed Points: A Holographic Check},''
  \href{http://dx.doi.org/10.1007/JHEP03(2011)041}{{\em JHEP} {\bf 1103} (2011)
   041}, \href{http://arxiv.org/abs/1011.5278}{{\tt arXiv:1011.5278 [hep-th]}}.

\bibitem{RRBenvenuti:2004dy}
S.~Benvenuti, S.~Franco, A.~Hanany, D.~Martelli, and J.~Sparks, ``{An infinite
  family of superconformal quiver gauge theories with Sasaki-Einstein duals},''
  {\em JHEP} {\bf 06} (2005)  064,
\href{http://arxiv.org/abs/hep-th/0411264}{{\tt arXiv:hep-th/0411264}}.

\bibitem{RREager:2012hx}
R.~Eager, J.~Schmude, and Y.~Tachikawa, ``{Superconformal Indices,
  Sasaki-Einstein Manifolds, and Cyclic Homologies},''
  \href{http://dx.doi.org/10.4310/ATMP.2014.v18.n1.a3}{{\em Adv. Theor. Math.
  Phys.} {\bf 18} (2014) no.~1, 129--175},
\href{http://arxiv.org/abs/1207.0573}{{\tt arXiv:1207.0573 [hep-th]}}.

\bibitem{RRGaiotto:2009gz}
D.~Gaiotto and J.~Maldacena, ``{The gravity duals of N=2 superconformal field
  theories},''
\href{http://arxiv.org/abs/arXiv:0904.4466}{{\tt arXiv:arXiv:0904.4466
  [hep-th]}}.

\bibitem{RRPeelaers:2014ima}
W.~Peelaers, ``{Higgs branch localization of $ \mathcal{N} $ = 1 theories on
  S$^{3}$ x S$^{1}$},'' \href{http://dx.doi.org/10.1007/JHEP08(2014)060}{{\em
  JHEP} {\bf 08} (2014)  060},
\href{http://arxiv.org/abs/1403.2711}{{\tt arXiv:1403.2711 [hep-th]}}.

\bibitem{RRYoshida:2014qwa}
Y.~Yoshida, ``{Factorization of 4d N=1 superconformal index},''
\href{http://arxiv.org/abs/1403.0891}{{\tt arXiv:1403.0891 [hep-th]}}.

\bibitem{RRNieri:2015yia}
F.~Nieri and S.~Pasquetti, ``{Factorisation and holomorphic blocks in 4d},''
  \href{http://dx.doi.org/10.1007/JHEP11(2015)155}{{\em JHEP} {\bf 11} (2015)
  155},
\href{http://arxiv.org/abs/1507.00261}{{\tt arXiv:1507.00261 [hep-th]}}.

\bibitem{RRPasquetti:2011fj}
S.~Pasquetti, ``{Factorisation of N = 2 Theories on the Squashed 3-Sphere},''
  \href{http://dx.doi.org/10.1007/JHEP04(2012)120}{{\em JHEP} {\bf 1204} (2012)
   120},
\href{http://arxiv.org/abs/1111.6905}{{\tt arXiv:1111.6905 [hep-th]}}.

\bibitem{RRBeem:2012mb}
C.~Beem, T.~Dimofte, and S.~Pasquetti, ``{Holomorphic Blocks in Three
  Dimensions},''
\href{http://arxiv.org/abs/1211.1986}{{\tt arXiv:1211.1986 [hep-th]}}.

\bibitem{RRBenini:2013yva}
F.~Benini and W.~Peelaers, ``{Higgs branch localization in three dimensions},''
  \href{http://dx.doi.org/10.1007/JHEP05(2014)030}{{\em JHEP} {\bf 05} (2014)
  030},
\href{http://arxiv.org/abs/1312.6078}{{\tt arXiv:1312.6078 [hep-th]}}.

\bibitem{RRBenini:2011nc}
F.~Benini, T.~Nishioka, and M.~Yamazaki, ``{4d Index to 3d Index and 2d
  TQFT},'' \href{http://arxiv.org/abs/1109.0283}{{\tt arXiv:1109.0283
  [hep-th]}}.

\bibitem{RRRazamat:2013opa}
S.~S. Razamat and B.~Willett, ``{Global Properties of Supersymmetric Theories
  and the Lens Space},''
  \href{http://dx.doi.org/10.1007/s00220-014-2111-0}{{\em Commun. Math. Phys.}
  {\bf 334} (2015) no.~2, 661--696},
\href{http://arxiv.org/abs/1307.4381}{{\tt arXiv:1307.4381 [hep-th]}}.

\bibitem{RRSpiridonov:2010em}
V.~P. Spiridonov, ``{Elliptic beta integrals and solvable models of statistical
  mechanics},'' {\em Contemp. Math.} {\bf 563} (2012)  181--211,
\href{http://arxiv.org/abs/1011.3798}{{\tt arXiv:1011.3798 [hep-th]}}.

\bibitem{RRYamazaki:2013nra}
M.~Yamazaki, ``{New Integrable Models from the Gauge/YBE Correspondence},''
  \href{http://dx.doi.org/10.1007/s10955-013-0884-8}{{\em J. Statist. Phys.}
  {\bf 154} (2014)  895},
\href{http://arxiv.org/abs/1307.1128}{{\tt arXiv:1307.1128 [hep-th]}}.

\bibitem{RRGaiotto:2014ina}
D.~Gaiotto and H.-C. Kim, ``{Surface defects and instanton partition
  functions},''
\href{http://arxiv.org/abs/1412.2781}{{\tt arXiv:1412.2781 [hep-th]}}.

\bibitem{RRBullimore:2014nla}
M.~Bullimore, M.~Fluder, L.~Hollands, and P.~Richmond, ``{The superconformal
  index and an elliptic algebra of surface defects},''
\href{http://arxiv.org/abs/1401.3379}{{\tt arXiv:1401.3379 [hep-th]}}.

\bibitem{RRAlday:2013rs}
L.~F. Alday, M.~Bullimore, and M.~Fluder, ``{On S-duality of the Superconformal
  Index on Lens Spaces and 2d TQFT},''
\href{http://arxiv.org/abs/1301.7486}{{\tt arXiv:1301.7486 [hep-th]}}.

\bibitem{RRRazamat:2013jxa}
S.~S. Razamat and M.~Yamazaki, ``{S-duality and the N=2 Lens Space Index},''
  \href{http://dx.doi.org/10.1007/JHEP10(2013)048}{{\em JHEP} {\bf 1310} (2013)
   048},
\href{http://arxiv.org/abs/1306.1543}{{\tt arXiv:1306.1543 [hep-th]}}.

\bibitem{Yagi:2015lha}
J.~Yagi, ``{Quiver gauge theories and integrable lattice models},''
  \href{http://dx.doi.org/10.1007/JHEP10(2015)065}{{\em JHEP} {\bf 10} (2015)
  065},
\href{http://arxiv.org/abs/1504.04055}{{\tt arXiv:1504.04055 [hep-th]}}.

\bibitem{Maruyoshi:2016caf}
K.~Maruyoshi and J.~Yagi, ``{Surface defects as transfer matrices},''
\href{http://arxiv.org/abs/1606.01041}{{\tt arXiv:1606.01041 [hep-th]}}.

\end{thebibliography}\endgroup

\fi
}

\newcommand{\documentfinishBBL}{
\addcontentsline{toc}{section}{References}
\ifx\ifLONG\undefined
\input{\IDvalue.separate.bbl}
\end{document}
\else
\input{\DIRvalue/\IDvalue.volume.bbl}
\fi
}

\ifx\ifLONG\undefined
\def\volcite#1{Contribution \cite{Contribution#1}}
\else 
\def\volcite#1{Chapter \ref{Chapter#1}}
\fi

\usepackage{xcolor}
\usepackage{array}

\usepackage{mathrsfs}
\usepackage{amsfonts}
\usepackage{amsmath}
\usepackage{yfonts}

\usepackage{dsfont}
\usepackage{bbm}
\usepackage{colonequals}
\usepackage{amscd}
\usepackage{skak}
\usepackage{graphicx}
\usepackage{hyperref}
\usepackage{fullpage}

\numberwithin{equation}{section}

\newcommand{\wt}{\widetilde}
\newcommand{\ep}{\epsilon}

\newcommand{\cA}{\mathcal{A}}
\newcommand{\cB}{\mathcal{B}}
\newcommand{\cC}{\mathcal{C}}
\newcommand{\cD}{\mathcal{D}}

\newcommand{\cI}{\mathcal{I}}

\newcommand{\cN}{\mathcal{N}}
\newcommand{\cO}{\mathcal{O}}

\newcommand{\cQ}{\mathcal{Q}}

\newcommand{\cS}{\mathcal{S}}

\newcommand{\BS}{\mathbb{S}}

\newcommand{\Tr}{{\rm Tr}}
\begin{document}
\thispagestyle{empty}
\documentheader
\else \def\workingdir{RastelliRazamat}
\chapterheader  \fi

\newcommand{\RRcheck}{\textbf{CHECK!}}

\newcommand{\RRbea}{\begin{eqnarray}}
\newcommand{\RReea}{\end{eqnarray}}
\newcommand{\RRbe}{\begin{equation}}
\newcommand{\RRee}{\end{equation}}

\newcommand{\RRben}{\begin{eqnarray}\displaystyle}
\newcommand{\RReen}{\end{eqnarray}}

\newcommand{\ND}{{\rm ND}}

\newcommand{\Ab}{\mathbb{A}}
\newcommand{\Cb}{\mathbb{C}}
\newcommand{\Fb}{\mathbb{F}}
\newcommand{\Hb}{\mathbb{H}}
\newcommand{\Kb}{\mathbb{K}}
\newcommand{\Mb}{\mathbb{M}}
\newcommand{\Nb}{\mathbb{N}}
\newcommand{\Ob}{\mathbb{O}}
\newcommand{\Qb}{\mathbb{Q}}
\newcommand{\Rb}{\mathbb{R}}
\newcommand{\Zb}{\mathbb{Z}}

\numberwithin{equation}{section}

\section{Introduction}

The technique of supersymmetric localization allows to compute 
the partition functions of several supersymmetric field theories on certain compact manifolds preserving some of the supersymmetry.
In favorable cases, the procedure of localization  reduces the computation of the infinite dimensional path integral to a finite dimensional integral or to a discrete  sum.  Many of the computable supersymmetric partition functions in dimension $d \leq 4$ are related to one another, see figure 1. The relations take two forms. First, different partition functions might be related by taking various limits of their parameters. 
For example, a partition function on a compact manifold can depend on  the relative size of different components -- sending that size to zero corresponds to computing a partition function of a theory in a lower dimension. Such limits are represented by solid lines in the figure. Second,  partition functions on compact manifolds can sometimes be computed by gluing together partition functions on non-compact manifolds with prescribed boundary conditions at infinity. Different patterns of gluing of the same non-compact partition functions can lead to two different compact partition functions.  For example,  both the $\BS^2\times \BS^1$ partition function 
(the three-dimensional supersymmetric index) and the  $\BS^3$ partition function are obtained by gluing partition functions on ${\mathbb C}\times \BS^1$.    Such relations are denoted by dashed lines in the figure. Some of the relations indicated in the figure are well studied while for others only some partial understanding is available.\footnote{This picture could be extended to a larger network of relations starting from higher dimensional theories --  the $\BS^4$ partition function \cite{RRPestun:2007rz} (see \volcite{HO}), notably absent in figure 1, would be part of
such an extended picture.}

\begin{figure}[ht]
\begin{center}
\includegraphics[scale=.71]{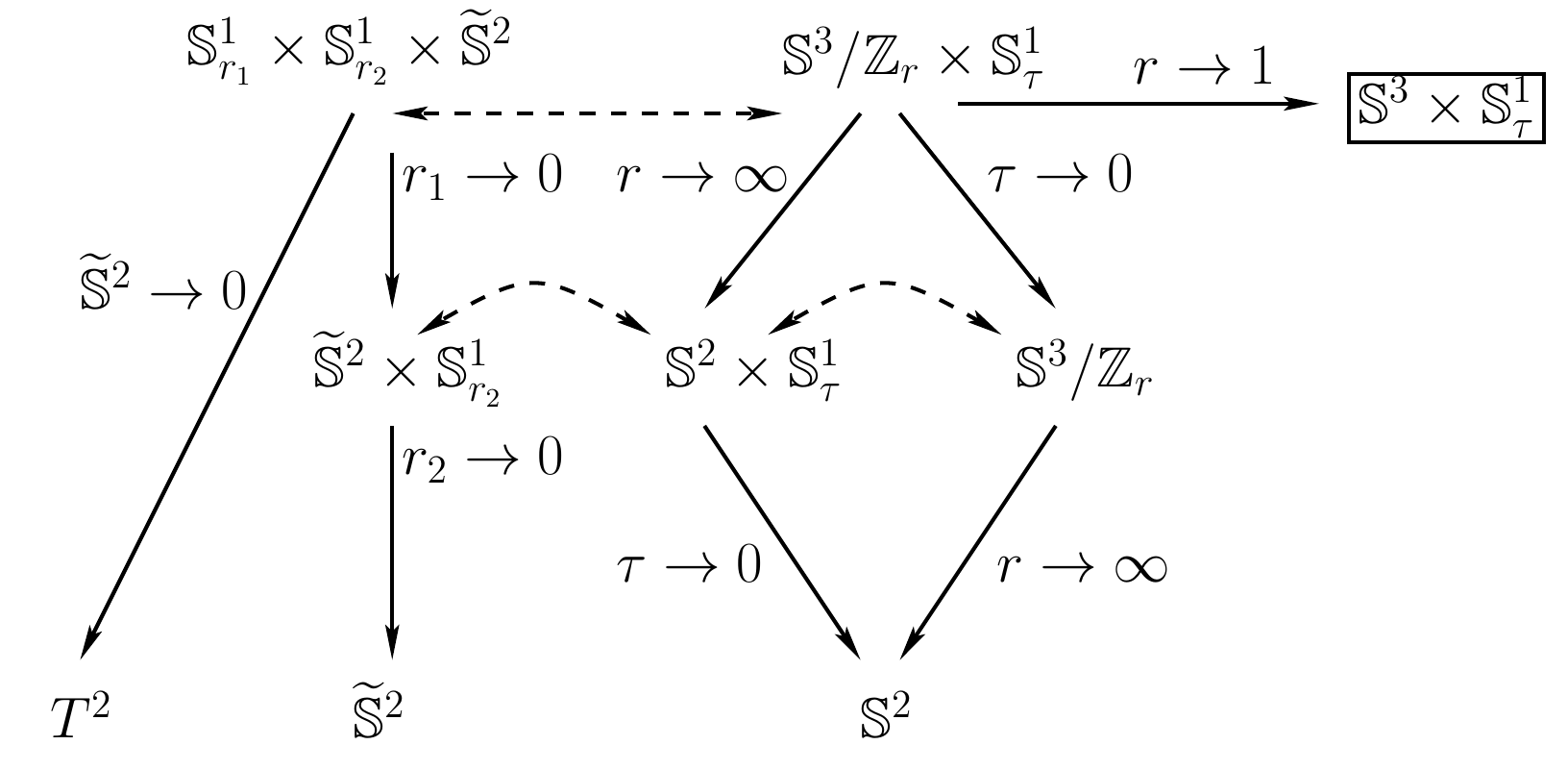}
\end{center}
\caption{Different supersymmetric partition functions in dimensions $4,3,2$ are related by limits of parameters (solid lines) and block decompositions (dashed lines). The
$\BS^3\times \BS^1$ partition function (also known as the four-dimensional index)
  is one of the simplest and most useful partition functions.}\label{RRlims}
\end{figure} 

The main focus of this review article will be the  $\BS^3\times \BS^1$ partition function, also known as the four-dimensional
 supersymmetric index, because it can be understood as the Witten index of the theory quantized on  $\BS^3\times {\mathbb R}$,  refined by  fugacities that keep track of the relevant quantum numbers.
 This is the simplest and arguably the most important observable in the network of partition functions shown in figure 1. For theories that admit a Lagrangian description, the four-dimensional index can be obtained by solving a simple counting problem: one enumerates (with signs)
 local gauge invariants operators built from elementary fields in the four dimensional theory,
 in the limit of vanishing coupling.  By contrast,  the supersymmetric index in other dimensions gets contributions from more complicated objects, such as instantons in five dimensions, monopoles in three dimensions, and local supersymmetric defects in two dimensions. The four-dimensional counting problem is efficiently encoded by a simple matrix integral, which could be equivalently obtained by applying the recipe of supersymmetric localization to the   $\BS^3\times \BS^1$ partition function. While the four-dimensional index is computationally simpler than other partition functions, its properties and the technology needed to extract physical information from it are of more universal applicability.

This review article is organized as follows. In section 2 we discuss the definition of the supersymmetric index and  the prescription to compute it in any Lagrangian theory. In section 3 properties of the index of theories built from chiral fields 
with superpotential interactions are reviewed. In section 4 we discuss basic properties of the index of gauge theories. 
In section 5 we review superconformal representation theory and the way different multiplets are encoded in the index. In particular we review how to extract easily the spectrum of relevant and exactly marginal deformations. In section 6 we discuss briefly some of the mathematical properties of indices. In particular we review symmetries of the index and identities between indices of different looking theories related by dualities. In section 7 we review different physically important limits of the index. Finally, in section 8 we mention several topics not covered in detail in this review.

\

\section{Definition of the index}
  
 There are two equivalent ways to define the supersymmetric index. It can be defined as the supersymmetric partition function on $\BS^3\times \BS^1$, which depends holomorphically
 on two complex structure moduli (conventionally denoted $p$ and $q$) and on holonomies for 
  background gauge fields coupling  to the flavor symmetries of the theory.
  Alternatively, it is given by an appropriately weighted trace over the states of the theory quantized on $\BS^3\times {\mathbb R}$. If the theory is conformal, one can use the state/operator map
  to interpret these states as local operators. Only in such cases it is appropriate to refer to the index as the {\it superconformal} index.
There are many important examples of ${\cal N}=1$ superconformal field theories that can be reached as infrared fixed points of RG flows starting 
 from  weakly-coupled Lagrangian theories.  One of the most useful properties of the index (most easily argued using its definition as a partition function)
 is its invariance under RG flow. This provides a powerful way to obtain the index of an IR fixed point, by performing a simple calculation in the UV.

\

\subsection{Index as a trace}

The  index of a $4d$ super{\it{conformal}} field theory is defined
as the Witten index of the theory in radial quantization.  Let ${\cal Q}$ be  one of the  Poincar\'e supercharges, and $\cQ^\dagger ={\cal  S}$
the conjugate conformal supercharge. Schematically, the index is defined as ~\cite{RRKinney:2005ej, RRRomelsberger:2005eg, RRRomelsberger:2007ec}
 \RRbe \label{RRbasicdef} {\mathcal I} (\mu_i)={\rm
Tr}\, (-1)^{F}\,e^{-\beta\, \delta}\, e^{-\mu_i {\cal M}_i}\, ,
\RRee
where the trace is over the Hilbert space  of the theory quantized on $\BS^3$,
$\delta \equiv \frac{1}{2} \{\cQ,\, \cQ^{\dagger}\}$,
  ${\cal M}_i$ are $\cQ$-closed
conserved charges and $\mu_i$ the associated chemical potentials. Since states with $\delta >0$
come in boson/fermion pairs, only the $\delta =0$ states contribute, and the index is independent of $\beta$.
There are  infinitely many states with $\delta =0$ --
this is true even  for a single  short irreducible representation of the superconformal algebra,
 because
some of the non-compact generators (some of the spacetime derivatives)
  have $\delta=0$. The introduction of the chemical
potentials $\mu_i$ serves both to regulate this divergence and to achieve a more
refined counting.

For  ${\cal N}=1$, the supercharges are $\{ {\cal Q}_\alpha \, , {\cal S}^\alpha \equiv {\cal Q}^{\dagger\, \alpha} \, , {\widetilde {\cal Q}_{\dot \alpha}} \, ,
\widetilde {\cal S}^{\dot \alpha} \equiv {\widetilde {\cal Q}^{\dagger \,\dot \alpha}} \}$,
where $\alpha = \pm$ and $\dot \alpha = \dot \pm$ are respectively $SU(2)_1$ and $SU(2)_2$ indices, with  $SU(2)_1 \times SU(2)_2 = Spin(4)$ the isometry group of the $\BS^3$.
The relevant anticommutators are
\RRbea
\{\cQ_\alpha, \, {\cal \cQ}^{\dagger\, \beta} \} & =& E+2M_\alpha^\beta+\frac{3}{2}r \\
\{\widetilde \cQ_{\dot \alpha}\,, \, {\widetilde {\cal \cQ}}^{\dagger \, \dot \beta} \} & =& E +2 \widetilde M_{\dot \alpha}^{\dot \beta}-\frac{3}{2}r \, ,
\RReea
where $E$ is the conformal Hamiltonian, $M_{\alpha}^\beta$ and  $\widetilde M_{\dot \alpha}^{\dot \beta}$  the $SU(2)_1$ and $SU(2)_2$ generators, and
$r$ the generator of the $U(1)_r$ R-symmetry. In our conventions, the $\cQ$s have $r=-1$ and $\widetilde Q$s have $r=+1$,
and of course the dagger operation  flips the sign of $r$.

One can define two inequivalent indices,
a ``left-handed'' index   $\cI^{{\tt L}}(t,y)$  and a ``right-handed'' index  $\cI^{{\tt R}}(t,y)$. For the left-handed index,
we pick say\footnote{Picking $\cQ \equiv\cQ_{+}$ would amount to the replacement $j_1 \leftrightarrow -j_1$, which is an equivalent choice because of $SU(2)_1$ symmetry. The same consideration applies to the right-handed index, which can be defined either choosing $\widetilde \cQ_{\dot -}$ or $\widetilde \cQ_{\dot +}$. } $\cQ \equiv \cQ_{-}$,
\RRbe
\label{RReq:defofindexL}
\cI^{{\tt L}}(p,q) \equiv {\rm Tr} \, (-1)^F p^{\frac13(E+j_1)+j_2}q^{\frac13(E+j_1)-j_2} = 
  {\rm Tr} \, (-1)^F p^{j_1+j_2-\frac12 r}q^{j_1-j_2-\frac12 r}\, ,\qquad
\delta=E-2j_1+\frac{3}{2}r \, ,
\RRee
where $j_1$ and $j_2$ are the Cartan generators of $SU(2)_1$ and $SU(2)_2$.
The two ways of writing the exponent of $t$ are equivalent since they differ by a $\cQ$-exact term. For the right-handed index,
we pick
say $\cQ \equiv \widetilde \cQ_{\dot -}$,
\RRbe
\label{RReq:defofindexR}
\cI^{{\tt R}}(p,q) \equiv {\rm Tr} \, (-1)^F p^{\frac13(E+j_2)+j_1}q^{\frac13(E+j_2)-j_1}  =   {\rm Tr} \, (-1)^F p^{j_1+j_2+\frac12 r}q^{j_2-j_1+\frac12 r}\, ,\qquad
\delta=E-2j_2-\frac{3}{2}r \,.
\RRee
One may also introduce chemical potentials for  global symmetries of the theory which commute with the supersymmetry algebra and thus conserve the index property of the trace. Such fugacities can be turned on for continuous and/or discrete symmetries as we will see in what follows.\footnote{One can consider additional generalizations of the index such as introduction  of charge conjugation~\cite{RRZwiebel:2011wa} to the trace but we will refrain from doing so here.}

If the theory is {\it not} conformal, and  is described instead by an RG flow from a free UV fixed point to an IR fixed point, one can still define the index from \eqref{RRbasicdef}, evaluating the trace over the local operators at the UV fixed point, but making  sure that the allowed
 symmetries are preserved along the flow. (In particular, the R-charge assignments must correspond to a non-anomalous R symmetry). 
Since the index is an RG invariant, this gives a recipe to evaluate the superconformal index of the IR fixed point.
At intermediate scales on the flow, the index is interpreted as the partition function on ${\mathbb S}^3\times {\mathbb S}^1$, or equivalently, as the trace over the states of the theory quantized on ${\mathbb S}^3$.

\

\subsection{Index as a partition function}

Alternatively, the index  can be defined as the supersymmetric partition function on $\BS^3\times \BS^1_\tau$.   As was argued in~\cite{RRFestuccia:2011ws}  (see also \cite{RRSen:1985ph} any ${\cal N}=1$ supersymmetric theory can be put in a supersymmetric way on $\BS^3\times \BS^1_\tau$ provided  it possesses anomaly free $U(1)_r$ R symmetry.  We refer to \volcite{DU}  for a detailed treatment and mention here only some of the salient points.

The $\BS^3\times \BS^1_\tau$ partition function depends holomorphically on the complex structure moduli $p$ and $q$, and on the holonomies associated to flavor symmetries. It does {\it not}
depend on gauge and superpotential couplings\footnote{But of course, the presence of a superpotential  restricts the possible R charge assignments.}, and is invariant under RG flow (\volcite{DU}).  
The partition function can be evaluated by localization techniques~\cite{RRClosset:2013sxa,RRAssel:2014paa}, and the result is the same matrix integral that we will obtain in the next subsection by enumeration of  gauge invariant operators.

The precise equivalence of the trace formula for the index and the computation of the partition function requires a bit of care. 
When computing the index using the trace formula we implicitly normalize it so the vacuum (assuming it is unique)  contributes $+1$ to the index.  In particular in the large radius limit, $\tau\to \infty$ the index computed as a counting problem receives only contributions from the vacua, and assuming there is a unique vacuum (preserving certain global symmetries) the index in the limit is 1. However, while computing the partition function in the large radius limit one finds a contribution coming from the Casimir energy of the theory,
\RRbe
\lim_{\tau\to\infty} {\cal Z}_{\BS^3\times\BS^1_\tau} \sim e^{-\tau E_{Casimir}}\,.
\RRee The trace formulation of the index and the partition function formulation thus differ by the multiplicative factor $e^{-\tau E_{\rm Casimir}}$. The Casimir energy can be computed from the trace formulation of the index~\cite{RRKim:2012ava,
RRArdehali:2015bla, RRAssel:2015nca, RRBobev:2015kza},
\RRbea
E_{\rm Casimir} & =&  -\lim_{\tau\to\infty} \frac{d}{d\tau} \log {\cal I}(p=e^{-\tau \omega_1},q=e^{-\tau \omega_2}, u_a=e^{i m_a \tau})\\
&=& \frac{2}{3} (a-c) (\omega_1 + \omega_2) + \frac{2}{27} (3c - 2a) \frac{(\omega_1 + \omega_2)^3}{\omega_1 \omega_2}\, ,
\RReea
where $a$ and $c$ are the Weyl anomaly coefficients.

\

\subsection{Computation of the index}

By the state/operator correspondence the computation of the index of a conformal gauge theory proceeds by listing all the possible operators we can construct from modes of the fields and projecting out gauge non invariant ones. The different modes of the fields are usually called ``letters'' and the operators are words constructed using this alphabet.
  
The   ``letters'' of an ${\cal N}=1$ chiral multiplet
are enumerated
in table \ref{RRtab1}. We assume that in the IR the $U(1)_r$ charge
of the lowest component of the multiplet $\phi$ is some arbitrary $r_{IR}=r$
(determined in a concrete theory by anomaly cancellation and in subtle cases $a$-maximization).
According to the prescription we have just reviewed,
the index receives contributions from the letters with $\delta_{UV}=0$,
and each letter contributes as $(-1)^F p^{j_1+j_2-\frac12r_{IR}}q^{j_1-j_2-\frac12r_{IR}}$ to the left-handed
index and as  $(-1)^Fp^{j_1+j_2+\frac12r_{IR}}q^{j_2-j_1+\frac12r_{IR}}$  to the right-handed index.
\begin{table}[htbp]
\begin{centering}
\begin{tabular}{|c||c|c|c|c|c||c|c||c|c|}
\hline Letters & $E_{UV}$ & $j_{1}$ & $j_{2}$ & $r_{UV}$ & $r_{IR}$
& $\delta_{UV}^{{\tt L}}$ & ${\cal I}^{\tt L}$ & $\delta_{UV}^{{\tt
R}}$ & ${ \cal I}^{\tt R}$ \tabularnewline \hline \hline $\phi$ &
$1$ & $0$ & $0$ & $\frac{2}{3}$ & $r$ & $2$ & $-$ & $0$ &
$(pq)^{\frac12r}$\tabularnewline \hline $\psi$ & $\frac{3}{2}$ &
$\pm\frac{1}{2}$ & $0$ & $-\frac{1}{3}$ & $r-1$ & $0^+,2^-$ &
$-(pq)^{\frac{2-r}2}$ & $2$ & $-$ \tabularnewline \hline $\partial \psi$ &
$\frac{5}{2}$ & $0$ & $\pm\frac{1}{2}$ & $-\frac{1}{3}$ & $r-1$ &
$2$ & $-$
 & $4^+,2^-$ & $-$ \tabularnewline \hline $\square\phi$ & $3$ & $0$ & $0$ &
$\frac{2}{3}$ & $r$ & $4$ & $-$ & $2$ & $-$ \tabularnewline \hline
\hline $\bar{\phi}$ & $1$ & $0$ & $0$ & $-\frac{2}{3}$ & $-r$ & $0$
& $(pq)^{\frac12r}$ & $2$ & $-$\tabularnewline \hline $\bar{\psi}$ &
$\frac{3}{2}$ & $0$ & $\pm\frac{1}{2}$ & $\frac{1}{3}$ & $-r+1$ &
$2$ & $-$ & $2^+,0^-$ & $-(pq)^{\frac{2-r}2}$\tabularnewline \hline
$\partial\bar{\psi}$ & $\frac{5}{2}$ & $\pm\frac{1}{2}$ & $0$ &
$\frac{1}{3}$ & $-r+1$ & $2^+,4^-$ & $-$ & $2$ & $-$ \tabularnewline
\hline $\square\bar{\phi}$ & $3$ & $0$ & $0$ & $-\frac{2}{3}$ & $-r$
& $2$ & $-$
 & $4$ & $-$\tabularnewline \hline\hline
$\partial_{\pm\pm}$ & $1$ & $\pm\frac{1}{2}$ & $\pm\frac{1}{2}$ &
$0$ & $0$ & $0^{\pm+},2^{\pm-}$ & $p,\,q$ &
$0^{+\pm},2^{-\pm}$ & $p,\,q$\tabularnewline \hline
\end{tabular}
\par\end{centering}
\caption{\label{RRtab1}The ``letters'' of an ${\cal N}=1$ chiral
multiplet and their contributions to the index. Here $\delta^{\tt L}=E-2j_{1}+\frac{3}{2}r_{UV}$ and
$\delta^{\tt R}_{UV}=E-2j_{2}-\frac{3}{2}r_{UV}$.  A priori
we have to take into account the free equations of motion $\partial \psi =0$ and $\Box \phi =0$,
which imply constraints on the possible words,
but we see that in this case equations of motions have $\delta_{UV} \neq 0$ so they do not change the index. Finally there
are two spacetime derivatives contributing to the index, and their multiple
action on the fields is responsible for the denominator of the index, $\frac1{(1-p)(1-q)} = \sum_{n,m=0}^\infty p^nq^m$.
}
\end{table}
To keep track of the gauge and flavor quantum numbers,
we introduce  characters. We assume that the chiral multiplet
transforms in the representation   ${\cal R}$ of the gauge $\times$ flavor group,
and denote by $\chi_{\cal R}(U, V)$,  $\chi_{\bar {\cal R}}(U, V)$ the characters of  ${\cal R}$ and
and of the conjugate representation $\bar {\cal R}$, with $U$ and $V$  gauge and flavor group matrices respectively.
All in all, the single-letter left- and right-handed indices for a chiral multiplet are \cite{RRDolan:2008qi}
\RRbea
 i_{\chi(r)}^{\tt L}(p,q,U, V) &  =&
\frac{(pq)^{\frac12r}\,\chi_{\bar{\mathcal R}}(U, V)-(p q)^{\frac{2-r}2}\,\chi_{\mathcal
R}(U, V)}{(1-p)(1-q)} \\  
\label{RRiC1}
  i_{\chi(r)}^{\tt R}(p,q,U, V) &  = &
\frac{(pq)^{\frac12r}\,\chi_{{\mathcal R}}(U, V)-(pq)^{\frac{2-r}2}\,\chi_{\bar {\mathcal
R}}(U,V)}{(1- p)(1- q)}\,. 
 \RReea
The denominators encode the action of the two spacetime derivatives with $\delta = 0$.
Note that the left-handed and right-handed indices differ by conjugation of the gauge and flavor quantum numbers.
As a basic consistency check \cite{RRRomelsberger:2007ec}, consider a single free massive chiral multiplet (no gauge or flavor indices).
In the UV, we neglect the mass deformation and as always $r_{UV} = \frac{2}{3}$. In the IR, the quadratic superpotential
implies $r_{IR}=1$, and one finds  $i_{r=1}^{\tt L} =  i_{r=1}^{\tt R} \equiv 0$. As expected, a massive superfield decouples and
does not contribute to the IR index.

Finding the contribution to the index of an ${\cal N}=1$ vector multiplet  is even easier,
since the $R$-charge of a vector superfield $W_\alpha$ is fixed at the canonical value $+1$
all along the flow.
For  both left- and the right-handed index,
the single-letter index of a vector multiplet is
~\cite{RRKinney:2005ej}
\RRbe \label{RRiV1}
i_{V}(p,q,U)  =
\frac{2pq-p-q}{(1-p)(1-q)}\,\chi_{adj}(U) \, . \RRee

Armed with the single-letter indices, the full index is obtained by enumerating all the
words and then projecting onto gauge-singlets
 by integrating over the Haar measure of the gauge group.
Schematically,
\begin{equation}\label{RRintegral}
{\cal I}(t, y, V)=\int[dU] \, \prod_k  \,  \, {\rm PE}[i_k (p, q, U, V)] \,,
\end{equation}
where $k$ labels the different supermultiplets,
and ${\rm PE}[i_k]$ is the plethystic exponential of the single-letter index of the $k$-th multiplet.
The pletyhstic exponential,
\RRbe
{\rm PE}[i_k(t, y, U, V)] \equiv \exp\left\{ \sum_{m=1}^{\infty}\frac{1}{m}i_k (p^{m}, q^m, V^m)\chi_{{\mathcal R}_{k}}(U^{m}, V^m)\right\}\,,
\RRee
implements the combinatorics of  symmetrization of the single letters, see {\it e.g.} \cite{RRBenvenuti:2006qr, RRAharony:2003sx}.
 As usual,  one can gauge fix the integral
over the gauge group and reduce it to an integral over the maximal torus, with the usual extra factor arising of van der Monde determinant.

The multi-letter contribution to the index of a chiral multiplet  (the plethystic exponential
of its single-letter index) can be elegantly written as a product of elliptic Gamma
functions~\cite{RRDolan:2008qi}.
For a chiral superfield in
the fundamental  representation $\Box$  of $SU(N_c)$, and with IR R-charge equal to $r$,
one has
\RRbea \label{RRchiralelliptic}
 {\rm PE}[i_r(p, q, U)]& \equiv& \prod_{i=1}^{N_c}\Gamma((p q)^{\frac12r}\,z_i;\,p,q),\qquad\\
\Gamma(z;p,q)& \equiv & \prod_{k,m=0}^\infty
\frac{1-p^{k+1}q^{m+1}/z}{1-p^{k}q^{m}\,z}\, .\nonumber  
\RReea
Here   $\{ z_k \}$,  $k=1, \dots N_c \}$ are complex numbers of unit modulus, obeying
$\prod_{k=1}^{N_c} z_k=1$, which
parametrize the Cartan subalgebras of $SU(N_c)$.

Similarly, the multi-letter contribution of a vector multiplet
in the adjoint of $SU(N)$ combines with the $SU(N)$ Haar measure to give
 the compact expression \cite{RRDolan:2008qi, RRGadde:2009kb}
\RRbe \label{RRvectorelliptic}
\frac{\kappa^{N-1}}{N!}\oint_{\mathbb{T}_{N-1}}\prod_{i=1}^{N-1}
\frac{dz_{i}}{2\pi i\,z_{i}}\,\prod_{k\neq
\ell}\frac{1}{\Gamma(z_k/z_\ell;p,q)}\,\dots\,. \RRee
The dots indicate that this is to be understood as a building block of the full matrix integral.
Here $\kappa$ is taken to be,
\RRbe
\kappa \equiv (p;p)(q;q) 
\RRee where $(a;b)\equiv\prod_{k=0}^\infty(1-ab^k)$. Note that $\kappa$ is the index of $U(1)$ free vector multiplet and we will sometimes denote $\kappa={\cal I}_V$.
 We will often
leave implicit the $q$ and $p$ dependence of the elliptic gamma
functions, $\Gamma(z; p,q) \to \Gamma(z)$. Also,  we will often use the shorthand notation
\RRbe
\Gamma(A z^{\pm1})\equiv \Gamma(A z)\Gamma(A z^{-1})\,.
\RRee

If the gauge group of the theory has abelian factors, one can turn on FI terms. On $\BS^3\times \BS^1_\tau$ such FI terms should be quantized~\cite{RRAharony:2013dha}. Indeed, on $\BS^3\times {\mathbb R}$ with sphere of radius $r_3$ the FI parameter $\zeta$ appears in the action as,
\RRbe
\zeta \int d^4x \sqrt{g}(D-\frac{2i}{r_3} A_4)\,,
\RRee where $A_4$ is the component of the gauge field along ${\mathbb R}$ and $D$ is the auxiliary field of the ${\cal N}=1$
vector multiplet. 
Upon compactification of ${\mathbb R}$ to $\BS^1_\tau$ we have to insure that this term is invariant under large gauge transformations, $A_4\to A_4+\frac1{\tau}$. Under such a transformation,
\RRbe
\zeta \int d^4x \sqrt{g}(D-\frac{2i}{r_3} A_4) \to \zeta \int d^4x \sqrt{g}(D-\frac{2i}{r_3} A_4) +
8\pi^3 i \zeta r_3^3\,, 
\RRee which implies that $\zeta=\frac1{4\pi^2 r_3^3} n$ with $n\in {\mathbb Z}$. The FI parameter for the $U(1)_u$ 
gauge factor will introduce the term $u^n$ in the matrix integral that computes the index.
\

The index does not depend on  any continuous coupling of the theory. However, the 
 functional form of the superpotential
restricts the possible global symmetries and hence the fugacities that the index can depend on. In turning on a certain set
of  fugacities, we are  computing the index for all possible choices of superpotentials consistent with the symmetries associated
to those fugacities.

\

\section{Index of sigma models}

We now turn to discuss basic properties of the index of some of  simplest ${\cal N}=1$ theories: sigma models built from chiral fields with no gauge interactions. 

\

\noindent $\bullet$ {\it Mass terms } --  Invariance along the RG flow is a basic property of the index. A simple implication is
  that
the index for a massive  theory with a single supersymmetric vacuum must be equal to $1$. 
Let's check this fact in the theory of two chiral fields with a superpotential mass term
\RRbe
W= m \, Q_a Q_b\,.
\RRee 
As the superpotential has R-charge two,
 the R-charges of the two fields satisfy \RRbe r_a+r_b=2 \,.\RRee Moreover there is one $U(1)$ symmetry under which the two fields are oppositely charged. Let us turn on a fugacity $u$ for this symmetry and assign charge $+1$ to field $a$. From our general rules, the index of this theory is
\RRbe
\Gamma((pq)^{\frac12r_a} u)\Gamma((pq)^{\frac12(2-r_a)}u^{-1})=
\prod_{i,j=0}^\infty \frac{1-(pq)^{1-\frac12r_a}p^i q^j u^{-1}}{1-(p q)^{\frac12 r_a}p^i q^j u}
\prod_{i,j=0}^\infty \frac{1-(p q)^{1-\frac12 (2-r_a)}p^i q^j u}{1-(pq)^{\frac12(2-r_a)}p^i q^j u^{-1}}=1\, ,
\RRee 
as expected.

\

\noindent $\bullet$  {\it F-term supersymmetry breaking} -- As another degenerate example, consider the theory of a chiral field with linear superpotential, $W=\eta Q$, the Polonyi model.
This model has no supersymmetric vacuum and thus breaks supersymmetry spontaneously. The field $Q$ has R-charge $2$ and is not charged under any global symmetry. The index is
\RRbe
\Gamma(pq)=0\, ,
\RRee  consistently with the absence of a supersymmetric vacuum. The vanishing the index can be traced to the presence
of a fermionic letter that contributes -1 (see Table 1): this mode should be interpreted as the Goldstino of supersymmetry breaking.
In general, models with spontaneous supersymmetry breaking of O'Raifeartaigh type will involve fields with R-charge two 
neutral under all global symmetries --  resulting in a vanishing index.

\

\noindent $\bullet$ {\it Runaway vacuum} -- We can consider a slight modification of the above model to restore the supersymmetric vacuum but at infinity in field space. We take
\RRbe
W=\eta Q+\frac12 \lambda Q^2 S\,.
\RRee The potential of this model has a minimum at zero as $S$ goes to infinity -- a runaway behavior. Indeed, 
the F-term equations read
\RRbe
\eta+ \lambda Q S=0,\qquad Q^2=0\,.
\RRee The vacuum is reached by taking the limit
\RRbe
Q\to 0,\qquad S\to \infty,\qquad  Q S =-\frac\eta\lambda\,.
\RRee 
The field $Q$ has R-charge $+2$ and contributes zero to the index (because of the fermionic zero mode mentioned above), while $S$ has R-charge $-2$ and contributes infinity, making the index of this model  ill-defined. 
The divergence in the index of the $S$ field can be traced to the existence of a bosonic zero mode, namely
$\partial_{-+}\partial_{++} \phi$, which contributes in the plethystic exponential with weight $1$ (see Table 1). 
 As we will soon discuss, divergences in the 
index signal the appearance of flat directions.
 In this example, the vacuum at infinity has a flat direction since the F-term equations are projective -- 
  it is  this flat direction that  gives rise to the divergent contribution.

\

\noindent $\bullet$ {\it Non-trivial chiral ring} -- Next, let us consider a superpotential of   the form $W=\lambda Q^{h+1}$ for some integer $h$. This model has a chiral ring relation $Q^h \sim 0$. The field $Q$ has R-charge $\frac2{h+1}$, it is not charged under any continuous global symmetries, 
but can carry charge under ${\mathbb Z}_{h+1}$. Let us denote by $g$ ($g^{h+1}=1$) the fugacity for ${\mathbb Z}_{h+1}$ and write the index of this model as
\RRbe
\Gamma((pq)^{\frac1{1+h}} g) = {\rm PE}[\frac{(pq)^{\frac1{1+h} }g-((pq)^{\frac{1}{1+h}}g)^h}{(1-p)(1-q)}]\,.
\RRee  Recall that the numerator in the plethystic exponential of a chiral field comes from the bosonic mode $\phi$ and a fermionic mode $\bar\psi$, while the denominator comes from the derivatives, $\partial_{\pm+}$.  Note that  $\bar\psi$ contributes to the index the $h$th power of the contribution of $\phi$ with an opposite sign. This implies  that the contribution of $\phi^h$ is cancelled by the contribution of $\bar \psi$, in accordance with the chiral ring relation discussed above.

\

\

\section{Index of gauge theories}\label{RRgaugeind}

\noindent $\bullet$ {\it D-term supersymmetry breaking} -- Let us first discuss the simplest gauge theory, $U(1)$ theory with 
an FI parameter $\zeta$, which as we discussed should be integer.  The index of this model is given by
\RRbe \label{Dterm}
\kappa \oint \frac{dz}{2\pi i z} z^\zeta =\kappa\, \delta_{\zeta,0}\,.
\RRee For non-zero FI parameter the index vanishes,  signalling  D-term supersymmetry breaking.
As we discussed in the previous section, pairs of chiral fields  with a mass term superpotential do not affect the index. The  index  (\ref{Dterm}) can then be interpreted as the index of a $U(1)$ gauge theory with any number of such pairs. Although the details of the dynamics of the model may depend on existence of such fields and on the relative values of the gauge coupling/FI term and masses, the index is always zero,  capturing only the fact that supersymmetry is broken.

\

\noindent $\bullet$ {\it IR duality} --  ${\cal N}=1$ gauge theories in four dimensions exhibit a variety of remarkable properties one of which is the ubiquity of IR dualities first discussed by Seiberg~\cite{RRSeiberg:1994pq}. A basic example  is ${\cal N}=1$ $SU(2)$ gauge theory with three flavors of fundamental and anti-fundamental quarks. This theory flows in the IR to a free theory in which is given by a sigma model of the collection of the mesonic and baryonic fields. The index of this gauge theory is given by
\RRbe\label{RRsugau}
{\cal I}_{gauge} = \kappa\oint \frac{dz}{4\pi i z} \frac1{\Gamma(z^{\pm2})} \prod_{i=1}^3
\Gamma((pq)^{\frac16} b u_i z^{\pm1})\Gamma((pq)^{\frac16} b^{-1} v_i z^{\pm1})\,.
\RRee 
Here $\prod_{i=1}^3u_i=\prod_{i=1}^3 v_i=1$, with these fugacities paramertizing the $SU(3)_u\times SU(3)_v$ flavor symmetry rotating the fundamental and anti-fundamental quarks, while $b$ parametrizes the baryonic $U(1)_b$. The distinction 
between fundamental and anti-fundamental matter here is artificial because of the pseudo-reality of the representations and is motivated by higher rank generalizations. In particular the $SU(3)_u\times SU(3)_v\times U(1)_b$ flavor symmetry enhances to $SU(6)_t$ with $\{t_i\}= \{ b u_i,\, b^{-1} v_i\}$. The index of the free mesons and baryons is given by
\RRbe 
{\cal I}_{sigma} = \prod_{i<j} \Gamma((pq)^{\frac13} t_i t_j)\,. 
\RRee If the index is to be independent of  the RG flow ${\cal I}_{gauge}$ should be equal to ${\cal I}_{sigma}$, which is indeed a proven mathematical fact. This identity is known as Spiridonov's beta function identity in math literature~\cite{spibrb}.  On the sigma model side we have fifteen chiral fields but the flavor symmetry has only rank five. The remaining symmetries rotating the chiral fields are broken by the superpotential which is is the Pfaffian of the antisymmetric matrix one can build from these fields. This superpotential is encoded in the index through the restriction of the fugacities to the ones of the $SU(6)_t$ symmetry.  

In evaluating the index, we have used the anomaly free R-charges for the quarks, $R=\frac13$. Mathematically, the anomaly free condition translates into a constraint on the arguments of the Gamma functions appearing in the numerator of the integrand. In this case we have, 
\RRbe
\prod_{i=1}^6 ((pq)^{\frac16}t_i)= p q\,.
\RRee Such constraints are called balancing conditions in the math literature~\cite{RRSpiridonov2}.

\

\noindent $\bullet$ {\it Higgsing/mass deformations} -- 
As discussed above, giving a mass to a pair of chiral fields trivializes their contribution to the index. If the theory has a dual IR description, the mass deformation 
corresponds to turning on a vacuum expectation value  that Higges the gauge symmetry on the other side of the duality. Let us discuss how this 
happens at the level of the index in a simple example.  We consider theory A to be an $SU(2)$ gauge theory with four flavors. This model has an $SU(4)_u \times SU(4)_v\times U(1)_b$ flavor symmetry. Is index is given 
\RRbe
{\cal I}_A({\bf u},{\bf v},b) = \kappa\oint \frac{dz}{4\pi i z} \frac1{\Gamma(z^{\pm2})} \prod_{i=1}^4 \Gamma((pq)^{\frac14} b u_i z^{\pm1})
\Gamma((pq)^{\frac14} b^{-1} v_i z^{\pm1})\, ,
\RRee where the fugacities satisfy the $SU(4)$ constraint
\RRbe\label{RRsucon}\prod_{i=1}^4 u_i=\prod_{i=1}^4 v_i =1\,.\RRee
 This model enjoys an IR duality. The Seiberg dual of it is a gauge theory with same rank and same charged matter content. However the charges of the quarks under global symmetries are different, they are in the conjugate representation
of the $SU(3)_u\times SU(3)_v$ flavor group.
 There are moreover gauge singlet fields having same charges as the mesons of the theory on side A and coupling to the mesons of the gauge theory on side B through a superpotential. The index of the theory on side B is
\RRbe\label{RRBside}
{\cal I}_B({\bf u}, {\bf v}, b) = {\cal I}_A ({\bf u}^{-1}, {\bf v}^{-1}, b) \, \prod_{i,j=1}^4 \Gamma((pq)^{\frac12} u_i v_j)\,.
\RRee The product over the Gamma functions is the product over the singlet fields. Thanks to an identity proved by Rains \cite{RRrains}, the indices on side A and side B coincide 
\RRbe
{\cal I}_A = {\cal I}_B\, ,
\RRee 
as expected from the duality. Again it was important here to use the anomaly free R-charges for the fields.

Let us now consider giving a mass to a pair of quarks on side A. This should give us the $SU(2)$ gauge theory with three flavors we discussed in the previous  bullet. We break the flavor symmetry from $SU(3)_u\times SU(3)_v$ down to $SU(2)_u\times SU(2)_v$. This breaking of symmetry through mass terms is encoded in the index by specializing the corresponding fugacities. For example, let us turn on a mass term $m Q_1\widetilde Q_1$. The weight of the mesonic operator  $Q_1\widetilde Q_1$ in the index before turning on the mass term is $(pq)^{\frac12} u_1 v_1$. After turning on the mass it should be $pq$ corresponding to R-charge $+2$ and no other charges.  Thus turning on the mass term in the index corresponds to specializing the fugacities to be
\RRbe\label{RRmasucon}
u_1 v_1 = (p q)^{\frac12} \,. 
\RRee  We now define $u_1=(p q)^{\frac14} a$, $v_1=(p q)^{\frac14} a^{-1}$, and find from \eqref{RRsucon},

\RRbe
\prod_{i=2}^4 u_i =(p q)^{-\frac14}a^{-1},\qquad \prod_{i=2}^4 v_i = (p q)^{-\frac14} a\,.
\RRee Redefining
\RRbe
u_i\equiv \tilde u_{i-1} (p q)^{-\frac1{12}} a^{-\frac13}\,,\qquad v_i\equiv \tilde v_{i-1} (p q)^{-\frac1{12}} a^{\frac13}\,,\qquad b = \tilde b a^{\frac13}\,,
\RRee  we obtain 
\RRbe 
\prod_{i=1}^3 \tilde u_i=\prod_{i=1}^3 \tilde v=1\,.
\RRee 
 After mass deformation, the index on side A becomes
\RRbe
{\cal I}_A \;\to\; \kappa \oint \frac{dz}{2\pi i z} \frac1{\Gamma(z^{\pm1})} \prod_{i=1}^3 
\Gamma((p q)^{\frac16} \tilde b \tilde u_i z^{\pm1})\Gamma((p q)^{\frac16} \tilde b^{-1} \tilde v_i z^{\pm1})\,,
\RRee which coincides with \eqref{RRsugau} as expected. 

Let us now discuss what happens on side B of the duality. Here the physics is more interesting. We gave a mass field to the meson $Q_1\widetilde Q_1$ on side A of the duality. On side B it maps to a singlet field, $M_{11}$, and thus the mass deformation adds a linear term to the superpotential. The superpotential involving the field $M_{11}$ is thus of the form
\RRbe
m M_{11} + q_1 \widetilde q_1 M_{11}\,, 
\RRee where $q_i$ and $\widetilde q_i$ are the quarks of the side B of the duality.  The F-term equation thus impose a vacuum expectation value for the meson $q_1\widetilde q_1$.  Turning such a vev Higgses the gauge $SU(2)$ gauge group and brings us to the sigma model of the previous bullet. Let us see what happens at the level of the index. The singlet $M_{11}$ contributes to the index as $\Gamma((pq)^{\frac12} u_1 v_1)$ and thus setting the fugacities to satisfy \eqref{RRmasucon} turns this into $\Gamma(pq)$ which is vanishing. 
 Let us analyze carefully what happens to the $SU(2)$ integral in \eqref{RRBside}. 
The integrand here has many poles in $z$. For example there are two poles coming from $\Gamma((p q)^{\frac14} b u_1 z^{\pm1})$ and two poles from $\Gamma((p q)^{\frac14} b^{-1} v_1 z^{\pm1})$ located at
\RRbe
z^{\pm1} =(p q)^{\frac14} b u_1\,,\qquad (p q)^{\frac14} b v_1\,.
\RRee Two of these poles are inside the $z$ integration contour and two are outside. 
Note then that if we specialize the fugacities to satisfy \eqref{RRmasucon} these four poles pinch the integration contour pairwise 
producing a divergence. The leading, divergent,  contribution to the integral  in the mass limit we consider thus comes only from two poles in the $z$ integral. These two poles are related by Weyl symmetry in the limit and thus give the same residues.
The divergence coming from the pinching is precisely canceled against the zero coming from the meson $M_{11}$ in the mass limit.
The index on side B in the limit is given then by
\RRbea
&&{\cal I}_B ({\bf u}^{-1}, {\bf v}^{-1}, b) \;\to\\
&&\; Res_{z\to (p q)^{\frac14} b u_1\,, u_1 v_1\to (p q)^{\frac12}} \left[
 \frac1{\Gamma(z^{\pm2})} \prod_{i=1}^4 \Gamma((pq)^{\frac14} b u_i z^{\pm1})
\Gamma((pq)^{\frac14} b^{-1} v_i z^{\pm1})\;\prod_{i,j=1}^4 \Gamma((pq)^{\frac12} u_i v_j)
\right]\nonumber\\
&&\qquad\qquad \to \prod_{i<j} \Gamma((p q)^{\frac13} \tilde t_i \tilde t_j)\,,\nonumber
\RReea where $\{\tilde t_i\}= \{\tilde b \tilde u_i,\tilde b^{-1}\tilde v_i\}$. We thus rederived the identity for the index
following from the duality of $SU(2)$  theory with there flavors to sigma model from the duality of $SU(2)$ theory with four flavors by following the RG flow triggered by mass term on one side of the duality and vev on the other side.

\

The general lesson to be learned here is that Higgsing gauge symmetries by vevs for gauge invariant operators 
manifests itself at the level of the index as reducing the number of integrals in the matrix model through the pinching procedure. In general a vev is possible when a flat direction opens up in the field space and this leads the index to have a pole. The index of the theory obtained in the IR of such an RG flow is given by the residue of the pole.

\

\noindent $\bullet$ {\it Spontaneously broken global symmetries} --  
We discussed spontaneous supersymmetry breaking above; here we will study a case of flavor symmetry breaking. 
The example we  consider  is $SU(2)$ gauge theory with two flavors, {\it i.e.}
two fundamental and two anti-fundamental quarks. This theory has an $SU(4)$ flavor symmetry 
at the classical level rotating the four quarks. However, at the quantum level the model can be described in terms of the  six
gauge singlet chiral fields $M_{ij}=Q_i Q_j$ with a quadratic constraint  ${\rm Pf}\, M =\Lambda^4$ where $\Lambda$ is the dynamical scale of the gauge theory. This dynamical superpotential breaks the $SU(4)$ symmetry down to $Sp(4)$.  

Let us see what happens here at the level of the index. The gauge theory at hand  can be obtained from the $SU(2)$ theory with three flavors we already considered by giving a mass to one of the flavors.   Let us denote the six quarks by $Q_i$ and 
rotate them with $SU(6)_t$ symmetry. We can turn on a mass term of the form $m Q_1 Q_2$. The theory with three flavors has an IR dual in terms of a sigma model and the analysis is simpler to perform on that side of the duality. Here we have a collection of fifteen singlet fields with a superpotential. The field $Q_1 Q_2$ is dual to singlet $M_{12}$. Turning on the mass term  the superpotential on the sigma model side involving field $M_{12}$ will become schematically
\RRbe
m M_{12} + M_{12} (M_{34} M_{56}+M_{36}M_{45}-M_{35}M_{46})\,.
\RRee In particular the F term coming from $M_{12}$  imposes the  constraint we discussed above,
\RRbe
m \sim M_{34} M_{56}+M_{36}M_{45}-M_{35}M_{46}\,.
\RRee  The weight of field $M_{12}$ before turning on the linear superpotential is $(p q)^{\frac13} t_1 t_2$ and after turning 
it on it becomes $p q$. Thus in the index we need to specialize the parameters as
\RRbe
t_1 t_2 = (p q)^{\frac23}\,.
\RRee We parametrize the fugacities as
\RRbe
t_1 =(p q)^{\frac13} a\,,\qquad  t_2 =(p q)^{\frac13} a^{-1}\,,\qquad t_{i>2} = (p q)^{-\frac16} \tilde t_{i-2}\,,\qquad
\prod_{i=1}^4 \tilde t_i = 1\,.
\RRee Fugacities $a$ and $\tilde t_i$ parametrize $u(1)_a\times su(3)_{\tilde t} =su(4)$ classical symmetry of the model.
Then after this specification the index of the sigma model becomes
\RRbe
{\cal I}_{sigma} \;\to \; \Gamma(p q) \, \prod_{i=1}^4 \Gamma( (p q)^{\frac12} a^{\pm1} \tilde t_i) \prod_{i<j} \Gamma(\tilde t_i\tilde t_j)\,.
\RRee This expression vanishes for generic values of $\tilde t_j$. In other words, if we insist on turning on fugacities for the classical $SU(4)$ symmetry the index vanishes indicating that there is no vacuum of the model having this symmetry. 
On the other hand let us further take $\tilde t_1 = \tilde t_2^{-1}\equiv c$. This also implies that $\tilde t_3= \tilde t_4^{-1}\equiv d$.  The symmetry we now parametrize is $su(2)_c\times su(2)_d \subset  sp(4)$.  After this specialization the index becomes
\RRbe
{\cal I}_{sigma} \;\to \; \Gamma(p q) \, \Gamma(1)^2  \Gamma( (p q)^{\frac12} a^{\pm1} c^{\pm1})\Gamma((p q)^{\frac12} a^{\pm1} d^{\pm1})  \Gamma(c^{\pm1} d^{\pm1}) =\Gamma(p q) \, \Gamma(1)^2   \Gamma(c^{\pm1} d^{\pm1})   \,.
\RRee Note that the fields charged under $U(1)_a$ can form mass terms and their contribution to the index trivializes.
Since $\Gamma(z)$  has a simple pole as $z\to 1$ and a simple zero as $z\to p q$, this expression diverges. We can thus summarize that unless we specialize the $SU(4)$ fugacities to parametrize an $Sp(4)$ subgroup the index vanishes and diverges otherwise.  The residue of the divergence  is given by
\RRbe
\Gamma(c^{\pm1} d^{\pm1})\,,
\RRee which is the  index of the collection of the chiral fields in any given quantum vacuum of the model. 

Let us consider the $SU(2)$ gauge theory with two flavors with the $Sp(4)$ flavor quantum symmetry, theory A, and some other theory with an $SU(2)_c$ flavor symmetry. which we will call theory B. Let us also assume that we can gauge in anomaly free fashion the  diagonal combination of the $SU(2)_c$ symmetry of theory B and an $SU(2)$ sub-group of the $Sp(4)$ symmetry of theory A. Note that at a generic point of the moduli space of theory A operator charged under $SU(2)_c$ obtains a vev. This Higgses the $SU(2)_c$ gauge group. Careful analysis reveals that the theory in the IR is identical to theory B with an addition of two singlet fields. We denote the index of theory B by ${\cal I}_B(c)$ where $c$ is fugacity for the $SU(2)_c$ symmetry. The index of the combined theory is then
\RRbe
{\cal I}(d,g)=\kappa^2 
\oint \frac{dc}{4\pi i c}\frac1{\Gamma(c^{\pm2})}
\oint_{\cal  C} 
\frac{dz}{4\pi i z} \frac1{\Gamma(z^{\pm2})} \Gamma(g\, c^{\pm1} z^{\pm1})\Gamma(g^{-1}\,d^{\pm1} z^{\pm1})
{\cal I}_B(c)\,.
\RRee 
One has to be careful here with the contour of integration since the poles of the index coming from the quarks of theory A sit on the unit circle. The contour can be obtained by carefully taking the mass limit from the theory with three flavors and we call it ${\cal C}$. This contour separates the sequences of poles these Gamma functions have converging to infinity and zero. 
Computation of this index reveals that it satisfies,

\RRbe\label{RRspiwar}
{\cal I}(d,g) = \Gamma(g^{\pm2}) \, {\cal I}_B(d)\,.
\RRee We have seen that the index of theory A vanishes except for a subset of fugacities where it diverges, and the above computation reveals that this index can be thought of as a delta function in the space of fugacities.  See~\cite{RRSpiridonov:2012ww} for more details.
The identity \eqref{RRspiwar} is known as an integral inversion formula of Spiridonov-Warnaar~\cite{RRspirinv}.

\

\section{Index spectroscopy}

The supersymmetric index contains useful information about the protected spectrum of the theory. The index counts (with signs) short multiplets up to the equivalence relation that sets to zero sets of short multiplets that may recombine into long ones. In general, it is not possible to deduce unambiguously from the index  the precise spectrum of short multiplets. However, for certain special multiplets corresponding to relevant and marginal operators, useful statements with a direct physical interpretation can be made. 
We will follow closely the discussion in~\cite{RRBeem:2012yn}.
 
\

A generic long multiplet $\cA_{r(j_{1},j_{2})}^{\Delta}$ of ${\cN}=1$ superconformal algebra is generated by the action of the four Poincar\'e supercharges $(\cQ_\alpha,\wt\cQ_{\dot{\alpha}})$ on a superconformal primary state, which by definition is annihilated by superconformal charges $(\cS_{\alpha},\wt\cS_{\dot\alpha})$. The multiplet is labeled by the charges $(\Delta,r,j_1,j_2)$ of the primary with respect to the dilatations, R-symmetry, and the two angular momenta respectively. The absence of negative norm states in the multiplet imposes certain inequalities on these quantum numbers,
\RRbea\label{RRunitarity}
\Delta~&&\geq~ 2-2\delta_{j_1,0}+2j_1-\frac 32 r,\\
\Delta~&&\geq~ 2-2\delta_{j_2,0}+2j_2+\frac 32 r, \\
\Delta~&&\notin (-\frac 32 r\,,\,2-\frac 32 r),\qquad\qquad\qquad\text{if}\;\;j_1=0\,,\\
\Delta~&&\notin (\frac 32 r\,,\,2+\frac 32 r)\,,\,\qquad\qquad\qquad\text{if}\;\;j_1=0\,,\\
\Delta~&&\geq~ 2+j_1+j_2\,,\qquad\qquad\qquad\text{if}\;\;j_1\neq0,\;\;j_2\neq0\,,\\
\Delta~&&\geq~ 1+j_1+j_2\,,\qquad\qquad\qquad\text{if}\;\;j_1=0\;\;\text{or}\;\;j_2=0~.
\RReea
When these inequalities are saturated, some combination of the Poincar\'e supercharges will annihilate the primary as well, resulting in a shortened multiplet. The relevant property of these short multiplets is that they must always saturate the unitarity bound in order to be free of negative normed states, and so their conformal dimension is fixed in terms of other quantum numbers and is protected against corrections as one changes the parameters of the theory.

The possible shortening conditions of the ${\cN}=1$ superconformal algebra are summarized in Table \ref{RRN1-shortening}. Note that  $\cD$ and $\bar{\cD}$ multiplets correspond to free fields and our general results below will not hold for them.

\begin{center}
\begin{table}[h!]
{\small
\begin{centering}
\begin{tabular}{|l|l|l|l|l|}
\hline
\multicolumn{4}{|c|}{Shortening Conditions} & Multiplet\\
\hline
$\cB$ & $\cQ_{\alpha}|r\rangle^{h.w.}=0$ & $j_1=0$ & $\Delta=-\frac{3}{2}r$ & $\cB_{r(0,j_2)}$\tabularnewline
\hline
$\bar{\cB}$ & $\bar{\cQ}_{\dot{\alpha}}|r\rangle^{h.w.}=0$ & $j_2=0$ & $\Delta=\frac{3}{2}r$ & $\bar{\cB}_{r(j_1,0)}$\tabularnewline
\hline
$\hat{\cB}$ & $\cB\cap\bar{\cB}$ & $j_1,j_2,r=0$ & $\Delta=0$ & $\hat{\cB}$\tabularnewline
\hline
$\cC$ & $\ep^{\alpha\beta}\cQ_{\beta}|r\rangle_{\alpha}^{h.w.}=0$ &  & $\Delta=2+2j_1-\frac{3}{2}r$ & $\cC_{r(j_1,j_2)}$\tabularnewline
 & $(\cQ)^{2}|r\rangle^{h.w.}=0$ for $j_1=0$ &  & $\Delta=2-\frac{3}{2}r$ & $\cC_{r(0,j_2)}$\tabularnewline
\hline
$\bar{\cC}$ & $\ep^{\dot{\alpha}\dot{\beta}}\bar{\cQ}_{\dot{\beta}}|r\rangle_{\dot{\alpha}}^{h.w.}=0$ &  & $\Delta=2+2j_2+\frac{3}{2}r$ & $\bar{\cC}_{r(j_1,j_2)}$\tabularnewline
 & $(\bar{\cQ})^{2}|r\rangle^{h.w.}=0$ for $j_2=0$ &  & $\Delta=2+\frac{3}{2}r$ & $\bar{\cC}_{r(j_1,0)}$\tabularnewline
\hline
$\hat{\cC}$ & $\cC\cap\bar{\cC}$ & $\frac{3}{2}r=(j_1-j_2)$ & $\Delta=2+j_1+j_2$ & $\hat{\cC}_{(j_1,j_2)}$\tabularnewline
\hline
$\cD$ & $\cB\cap\bar{\cC}$ & $j_1=0,-\frac{3}{2}r=j_2+1$ & $\Delta=-\frac{3}{2}r=1+j_2$ & $\cD_{(0,j_2)}$\tabularnewline
\hline
$\bar{\cD}$ & $\bar{\cB}\cap\cC$ & $j_2=0,\frac{3}{2}r=j_1+1$ & $\Delta=\frac{3}{2}r=1+j_1$ & $\bar{\cD}_{(j_1,0)}$\tabularnewline
\hline
\end{tabular}
\par\end{centering}
} \caption{\label{RRN1-shortening}Shortening conditions for the $SU(2,2\,|\,1)$ superconformal algebra.}
\end{table}
\par\end{center}\vspace{-15pt}
 
If the charges of a collection of short multiplets obey certain relations, they can combine to form a long multiplet which is no longer protected. Alternatively, one can understand this recombination in reverse, as a long multiplet decomposing into a collection short multiplets as the conformal dimension of its primary hits the BPS bound. This phenomenon plays a crucial role in extracting spectral information about an SCFT from its index because the index counts short multiplets of the theory \emph{up to recombination}. The collective contributions to the index from short multiplets that can recombine vanishes. The recombination equations for ${\cN}=1$ superconformal algebra are as follows:
\RRben\label{RRrecombination}
\cA_{r(j_{1},j_{2})}^{2+2j_{1}-\frac{3}{2}r} & \longrightarrow & \cC_{r(j_{1},j_{2})}\oplus\cC_{r-1(j_{1}-\frac{1}{2},j_{2})}~,\notag\\
\cA_{r(j_{1},j_{2})}^{2+2j_{2}+\frac{3}{2}r} & \longrightarrow & \bar{\cC}_{r(j_{1},j_{2})}\oplus\bar{\cC}_{r+1(j_{1},j_{2}-\frac{1}{2})}~,\\
\cA_{\frac{2}{3}(j_{1}-j_{2})(j_{1},j_{2})}^{2+j_{1}+j_{2}} & \longrightarrow & \hat{\cC}_{(j_{1},j_{2})}\oplus\cC_{\frac{2}{3}(j_{1}-j_{2})-1(j_{1}-\frac{1}{2},j_{2})}\oplus\bar{\cC}_{\frac{2}{3}(j_{1}-j_{2})+1(j_{1},j_{2}-\frac{1}{2})}~.\notag
\RReen
$\cB$ multiplets can be formally treated  as a special case of $\cC$ multiplets with unphysical spin quantum numbers,
\RRbe
\cB_{r(0,j_2)}=:\cC_{r+1(-\frac 12,j_2)}~,\qquad\qquad\bar{\cB}_{r(j_{1},0)}=:\bar{\cC}_{r-1(j_1,-\frac 12)}~.
\RRee
Thus the discussion can be phrased entirely in terms of ${\cC}$ type multiplets. 

\

An important  example of recombination  is for the long multiplet $\cA_{0(0,0)}^{2+-\frac32r}$ as $r\to0$. The multiplet hits the BPS bound and splits into three short multiplets according to the third rule in \eqref{RRrecombination},
\RRbe
\cA_{0(0,0)}^{2}\longrightarrow\hat{\cC}_{(0,0)}\oplus\cC_{-1(-\frac{1}{2},0)}\oplus\bar{{\cC}}_{1(0,-\frac{1}{2})}=\hat{{\cC}}_{(0,0)}\oplus({\cB}_{-2(0,0)}\oplus \bar{\cB}_{2(0,0)})
\RRee
The multiplet $\hat{\cC}_{(0,0)}$ contains a conserved current, while the multiplet $\cB_{-2(0,0)}$ contains a chiral primary $\cO$ of dimension three and an associated marginal F-term deformation $\int d^{2}\theta\,\cO$. The recombination described above demonstrates the fact that a marginal operator can fail to be exactly marginal if and only if it combines with a conserved current corresponding to a broken global symmetry. This particular recombination and its implications for the space of exactly marginal deformations of an SCFT has been studied in detail in \cite{RRGreen:2010da}.

\

The $\cC$ ($\bar{\cC}$) multiplets contribute only to the left-handed index (right-handed index), while $\hat\cC$ multiplets contribute to both. We restrict our attention to $\cI^{{\tt L}}$ and treat $\hat \cC$ as a special case of $\cC$ with $r=\frac{2}{3}(j_{1}-j_{2})$. The recombination rules allow us to define equivalence classes of short representations which make identical contributions to the index,
\RRbea
[\tilde r,j_2]_+&&\equiv\left\{\cC_{r(j_1,j_2)}\;|\;2j_1-r=\tilde r,\;\; j_1\in {\mathbb Z}_{\geq 0}\right\}\,,\nonumber\\
{[\tilde r,j_{2}]_{-} }&&\equiv\left\{\cC_{r(j_1,j_2)}\;|\;2j_1-r=\tilde r,\;\; j_1\in -\frac1{2}+{\mathbb Z}_{\geq 0}\right\}\,.
\RReea
For a $\cB$ type multiplet, the unitarity bounds of Equation \eqref{RRunitarity} imply that $\tilde r \geq -\frac 43+\frac 23 j_2$, while for a $\cC$ multiplet they imply $\tilde r \geq \frac 43 j_1 +\frac 23 j_2$. Consequently, there are a finite number of representations in a fixed equivalence class --- for fixed $\tilde r$, there is an upper limit on $j_1$ such that these bounds can be satisfied.

The contribution to the left-handed superconformal index from any short multiplet in a given class is given by

\RRbe\label{RRmultipletindex}
\cI_{[\tilde{r},j_{2}]_{+}}^{{\tt L}}=-\cI_{[\tilde{r},j_{2}]_{-}}^{{\tt L}}=(-1)^{2j_{2}+1}\frac{(p q)^{\frac12(\tilde{r}+2)}\chi_{j_{2}}(p/q)}{(1-p)(1-q)}~.
\RRee
We define the \emph{net degeneracy} for a given choice of $(\tilde r, j_2)$,
\RRbe
\ND[\tilde{r},j_{2}]:=\#\;[\tilde{r},j_{2}]_{+}-\#\;[\tilde{r},j_{2}]_{-}~,
\RRee
and the extractable content of the superconformal index is encapsulated in precisely the integers $\ND[\tilde r,j_2]$. If the index of an $\cN=1$ SCFT is known, the net degeneracies can be systematically extracted by means of a \emph{sieve algorithm} (see for example~\cite{RRBeem:2012yn}).
 The most precise information about actual operators we can extract from the index comes from the equivalence classes with a small number of representatives. 
 
The optimal case is the chiral primary operators that lie in multiplets $\cB_{r(0,j_2)}$ and have $-2-\tfrac23 j_2<r\leq-\tfrac 23-\tfrac 23 j_2$. These have $\tilde r \in [-\tfrac 43+\tfrac 23j_2,\tfrac 23j_2)$, and they are the \emph{only representatives} of the equivalence class $[\tilde r,0]_-$ for this range of $\tilde r$. Furthermore, there are no unitary representations in the corresponding class $[\tilde r,0]_+$. Consequently, we can read off the exact number of such operators from the superconformal index. Specializing to $j_2=0$, these are precisely the relevant deformations of the SCFT. The number of such deformations is simply the coefficient of $(p q)^{-\frac12r}(p/q)^{0}$ in the index after subtracting out any non-trivial $SU(2)_2$ characters at the same power of $p q$.

The next best case is for $\tilde r\in[\tfrac 23j_2,\tfrac 23+\tfrac 23j_2)$. Both $[\tilde r,j_2]_+$ and $[\tilde r,j_2]_-$ have only a single representative in this range, and so the index computes the difference in the number of such operators. For $j_2=\tilde r=0$ in particular, the representatives are $\hat{\cC}_{(0,0)}$ and $\cB_{-2(0,0)}$, respectively. The cancellation between these multiplets corresponds to precisely the recombination described in the example above, and we see that the index computes
\RRbe\begin{split}
\ND[0,0]  &= \#\ \cB_{-2(0,0)}-\#\ \hat{\cC}_{(0,0)}\cr
 & =  \#\:\mbox{marginal operators}-\mbox{\# }\mbox{conserved currents}~.
\end{split}\RRee
If all global flavor symmetries are broken at a generic point on the conformal manifold, then this net degeneracy will precisely capture the actual dimension of that conformal manifold. However, not all recombinations of the type discussed in the example necessarily take place, and in this case one must account for conserved currents in extracting the dimension of the conformal manifold. Again, this net degeneracy is easily computed by expanding the index to order $p q$ and subtracting out all nontrivial characters for $SU(2)_2$.

For $\tilde r\geq\tfrac 23$, there will be several representatives that are indistinguishable to the index, and the cancellations among them do not correspond to any obvious physical phenomenon such as symmetry breaking. Thus, the most immediate spectroscopic use of the index is the analysis of relevant and marginal operators at a fixed point.

\

\subsection{An example}

As an example we discuss $SU(N)$ ${\cal N}=4$ SYM. In ${\cal N}=1$ notation we have here three adjoint chiral fields, $\Phi_j$, with R-charge $\frac23$ rotated by $SU(3)_t$ global symmetry. The superconformal R-charge is that of a free field since the conformal manifold passes through the free point. The index is given by
\RRbe
{\cal I}_N(t,p,q)= \frac1{N!}\kappa^{N-1}\oint \prod_{j=1}^{N-1}\frac{dz_j}{2\pi i z_j} \prod_{j\neq k}\frac{\Gamma((p q)^{\frac13} t_1 z_j/z_k) \Gamma((p q)^{\frac13} t_2 z_j/z_k) 
\Gamma((p q)^{\frac13} \frac1{t_1 t_2}z_j/z_k) }
{\Gamma(z_j/z_k) }\,.
\RRee For $N>2$, the first few terms in the $p$, $q$ expansion are 
\RRbe\label{RRexpana}
{\cal I}_N(t,p,q) =1+ {\bf 6}_t (p q)^{\frac23}+ {\bf 3}_t(p+q) (p q)^{\frac13} +(1+{\bf 10}_t-{\bf 8}_t) pq+\cdots\,.
\RRee Following the general prescription of this section we read off the relevant operators to be ${\bf 6}_t$ which are the six quadratic operators $\Phi_{(j}\Phi_{k)}$. We have also operators charged under $j_2$ at order $(p+q) (p q)^{\frac13}$
which do not correspond to relevant operators. At order $p q$ we have the marginal operators. The contribution here is
$1+{\bf 10}_t-{\bf 8}_t$. The generators of the global symmetry form the ${\bf 8}_t$ which is subtracted the marginal operators are the gauge coupling and the ${\bf 10}_t$ symmetric cubic combinations of the adjoint chirals. At a generic point on the conformal manifold the $SU(3)_t$ symmetry is broken and the dimension of it is $1+10-8=3$ as expected. 
These exactly marginal deformations are the gauge coupling, the $\beta$ deformation (adding ${\rm Tr}\Phi_1\{\Phi_2,\,\Phi_3\}$ to superpotential), and the $\gamma$ deformation (adding also ${\rm Tr}(\Phi_1^3+\Phi_2^3+\Phi_3^3))$. 

The case of $N=2$ is special and there the expansion of the index coincides with~\eqref{RRexpana} except that ${\bf 10}_t$ term is missing. Here the conformal manifold is actually only one dimensional and corresponds to the gauge coupling. 
On any point of this manifold the $SU(3)_t$ symmetry is unbroken consistently with the index.  The reason here two directions are missing is that a general marginal superpotential cubic in the chiral fields can be decomposed as a sum of two terms, in one of which the gauge indices are contracted with $\epsilon_{abc}$ and the other with $d_{abc}= {\rm Tr}\,T_a\{T_b,\,T_c\}$. The latter structure is non zero only for $N>2$.

\

\section{Dualities and Identities}

Perhaps the most important application of the supersymmetric index as a test of non-perturbative dualities. 
Since the index is an RG invariant quantity and does not depend on the marginal couplings, it should be the same when computed for two theories flowing to the same fixed point or two different descriptions of the same conformal theory. 
Physical dualities translates into  mathematical
identities between elliptic hypergeometric integrals. Such identities are very non-trivial and give the strongest checks to date of many dualities.
In several cases, these identities have already appeared in the mathematical literature, but in many others they are new -- they are undoubtedly true since they can be checked
to very high orders in a series expansion, but a rigorous proof is still lacking.

\

\subsection{Symmetries and transformations of the index}

Before discussing relations between indices of dual theories, it is useful to pause and consider
the symmetry properties of the index of a single theory. The index is a function ${\cal I}(a_1,a_2,\cdots,a_s, p, q)$. The parameters $a_i$ are fugacities for $U(1)$ global symmetries forming the maximal torus of  the (possibly non-abelian) global symmetry.  If the symmetry enhances to a non-abelian symmetry the index should be invariant under the action of the Weyl group acting on the fugacities. For example, if the $a_i$'s parametrize an $SU(s+1)$ symmetry, the index should be invariant under  permutations of the $a_i$'s and under the transformation of any of the $a_j$ as $a_j\to \frac1{a_1a_2\cdots a_s}$.
 
We can also ask the converse question: what happens if the index is invariant under the action of certain discrete group ${\cal W}$ on the flavor fugacities?  There are two interesting physical possibilities. First, it might be that the flavor symmetry enhances to a non-abelian group such that ${\cal W}$ serves as its Weyl group. A second physical possibility
is that such a discrete symmetry signals  self-duality of the theory. An example is $SU(2)$ ${\cal N}=2$ SYM with four flavors. Here the flavour group (in ${\cal N}=2$ language) is rank four, and let us parametrize it by four fugacities $a_i$. In ${\cal N}=1$ language the index is given  by
\RRbe
{\cal I}(a_1,a_2,a_3,a_4)=\frac{\kappa\Gamma((p q)^{\frac13}t^{-2})}{2}\oint \frac{dz}{2\pi i z}\frac{\Gamma((p q)^{\frac13}t^{-2}z^{\pm2})}{\Gamma(z^{\pm2})} \Gamma((p q)^{\frac13}t a_1^{\pm1}a_2^{\pm1}z^{\pm1}) \Gamma((p q)^{\frac13}t a_3^{\pm1}a_4^{\pm1}z^{\pm1}) \,.
\RRee Here $t$ is fugacity for a $U(1)$ symmetry related to the bigger R-symmetry of ${\cal N}=2$. The flavor symmetry here enhance to $SO(8)$ and the index is manifestly invariant under the Weyl group of $SO(8)$. This group is generated by $a_i\to a_i^{-1}$ and by $a_1\leftrightarrow a_2$ , $a_3\leftrightarrow a_4$. However, the index is also invariant under exchanging $a_1$ and $a_3$. This is not part of $SO(8)$ Weyl symmetry and is not manifest in the integral above. This discrete symmetry is the manifestation of the self S-duality (or rather triality) that the theory enjoys. This is a strong/weak type of duality relating the same theory with different values of coupling. This invariance property of the index was proven in~\cite{RRFocco}.
  In fact the full discrete symmetry of the index, the one coming from Weyl of $SO(8)$ and the one coming from the duality, is the Weyl group of $F_4$. We are not aware of a physical interpretation
  for the full $F_4$ symmetry -- it would be nice to figure out whether there is any.
  
Another similar example is that of ${\cal N}=1$ $SU(2)$ theory with four flavors, {\it i.e.} the same theory as above but without the adjoint chiral field.  The theory has flavor symmetry of rank seven, the $SU(8)$ symmetry rotating the different matter fields. This theory enjoys Seiberg-duality as we already discussed, but in fact there are many more dualities as discussed in~\cite{RRSpiridonov:2008zr}. This theory in fact has $72$ dual descriptions. The different descriptions correspond to the action of the Weyl group of $E_7$ on the fugacities. In the different duality frames the gauge structure is the same as in the original one but there are additional singlet fields and superpotentials. It was argued in~\cite{RRDimofte:2012pd} that taking two copies of this theory coupled through a quartic superpotential the theory is exactly self-dual and that there should be a point on the conformal manifold of this theory where   the flavor symmetry is actually enhanced to $E_7$.

\

We can also ask whether there are interesting properties of the index involving manipulations of both the flavor fugacities and the superconformal fugacities $p$ and $q$.
A simple example is as follows. One can consider assigning different anomaly free R-charges to the fields by mixing a given R-symmetry with the flavor symmetry. For example give a flavor symmetry $U(1)_a$ we can redefine the R-symmetry to be $R\to R+s q_a$ with $q_a$ being the charge under $U(1)_a$. At the level of the index this transformation corresponds to 
\RRbe
R\to R+s q_a\;\Rightarrow\; {\cal I}(a,p,q) \to {\cal I}( (p q)^{\frac{s}2} a ,p,q)\,.
\RRee

Let us consider shifting flavor fugacity $a$ to $q^{\widetilde s} p^s a$. When $s$ and $\widetilde s$ are the same this is just a redefinition of the R-charge. From the definition of the index the shift in $a$ amounts to
\RRbe
{\cal I}= \Tr (-1)^F p^{j_1+j_2+\frac{r}2}q^{j_2-j_1+\frac{r}2} a^{q_a} \to\qquad\quad\;\;
\Tr (-1)^F p^{j_1+j_2+\frac{r}2+s q_a}q^{j_2-j_1+\frac{r}2+\widetilde s q_a} a^{q_a} \,.
\RRee To interpret this expression as an index we can redefine
\RRbe
\hat r= r+(s+\widetilde s) q_a\,, \qquad\quad \hat j_1 =j_1+\frac{s-\widetilde s}2 q_a\,.
\RRee In particular for $s\neq \widetilde s$ this breaks Lorentz symmetry and does not make sense as a pure $4d$ index.
However, such a transformation might make sense as an index of a coupled 4d-2d system.    
A simple example is the following important identity of the index of a chiral field,
\RRbe\label{RRindtt}
{\cal I}^{(R)}(p a) =\Gamma((p q)^{\frac{R}2}p a) = \theta((p q)^{\frac{R}2} a;q) \, {\cal I}^{(R)}(a)\,.
\RRee The index on the right-hand side can be interpreted as an index of chiral field in four dimensions coupled to a Fermi $(0,2)$ multiplet in two dimensions.  Similarly we have
\RRbe\label{RRindto}
{\cal I}^{(R)}(p^{-1} a) =\Gamma((p q)^{\frac{R}2}p^{-1} a) = \frac1{\theta((p q)^{\frac{R}2}p^{-1} a;q)} \, {\cal I}^{(R)}(a)\,.
\RRee Here the right hand side is a chiral field in four dimensions coupled to a chiral $(0,2)$ field in two dimension.
Such a transformation of the index will become important while discussing indices in presence of surface defects~\cite{RRGaiotto:2012xa,RR2014arXiv1412.2781G,RR2014JHEP...03..080G} and 
we will comment on this more in what follows.

\

\subsection{${\cal N}=4$ dualities}

A basic example of a duality implying a non-trivial mathematical identity is the S-duality between $SO(2n+1)$ ${\cal N}=4$ SYM and $USp(2n)$ ${\cal N}=4$ SYM.
 We use an ${\cal N}=1$ language with the three adjoint chiral multiplets having R-charge $\frac23$. 
Then the index of the  $SO(2n+1)$ model is given by
\RRbe
{\cal I}_{so} = \kappa^n \prod_{i=1}^3\Gamma((pq)^{\frac13}t_i)\frac1{2^nn!}
\oint \prod_{i=1}^n\frac{dz_i}{2\pi i z_i} \prod_{i<k}\frac{\prod_{j=1}^3\Gamma((pq)^{\frac13} t_j z_i^{\pm1}z_k^{\pm1})}
{\Gamma(z_i^{\pm1}z_k^{\pm1})}\prod_{i=1}^n\frac{\prod_{j=1}^3\Gamma((pq)^{\frac13} t_j z_i^{\pm1})}{\Gamma(z_i^{\pm1})}\, ,
\RRee while for the $USp(2n)$ model we get
\RRbe
{\cal I}_{sp} = \kappa^n \prod_{i=1}^3\Gamma((pq)^{\frac13}t_i)\frac1{2^nn!}
\oint \prod_{i=1}^n\frac{dz_i}{2\pi i z_i} \prod_{i<k}\frac{\prod_{j=1}^3\Gamma((pq)^{\frac13} t_j z_i^{\pm1}z_k^{\pm1})}
{\Gamma(z_i^{\pm1}z_k^{\pm1})}\prod_{i=1}^n\frac{\prod_{j=1}^3\Gamma((pq)^{\frac13} t_j z_i^{\pm2})}{\Gamma(z_i^{\pm2})}\,.
\RRee
Fugacities $t_j$ parametrize $SU(3)_t$ symmetry rotating the three adjoint chirals. We have decomposed the $SU(4)$ R-symmetry of ${\cal N}=4$ to $U(1)$ R-symmetry of ${\cal N}=1$ and $SU(3)_t$.
 For $n=1$ and $n=2$  the $SO(2n+1)$ and $USp(2n)$ algebras are isomorphic\footnote{The global form of the gauge group is inessential here --
  the spectrum of local gauge-invariant operators captured by the index depends only on the gauge algebra.} 
and there is a simple change of integration variables making the two expressions above manifestly the same.\footnote{The two root systems define  dual lattices in $n$ dimensions. In $n=1,2$ there is a linear transformation taking one into the other (line dual to line, and square dual to square), while for $n>2$ there is not.} 
For $n>2$, one can check that the two expressions coincide to very high orders in a series expansion in fugacities, but
no  proof is available yet except in certain degeneration limits~\cite{RRSpiridonov:2010qv}.
 
 \

\subsection{Seiberg dualities}

Seiberg dualities are the basic examples of IR dualities -- two theories flowing to the same fixed point. The simplest example 
is of an $SU(N)$ gauge theory with $N_f$ on side A
being equivalent to $SU(N_f-N)$ gauge theory with $N_f$ flavors, conjugate representation of the flavor group, and a bunch of gauge singlet fields dual to the mesons of side A on side B.  The index of side A is given by
\RRbe
{\cal I}_N^{N_f}({\bf u},{\bf v},b)=\kappa^{N-1}\frac1{N!}\oint \prod_{i=1}^{N-1} \frac{dz_i}{2\pi i z_i}
\prod_{i\neq j} \frac1{\Gamma(z_i/z_j)} \prod_{i=1}^{N_f}\prod_{j=1}^N \Gamma((p q)^{\frac{N_f-N}{2N_f}}b u_i z_j)
\Gamma((p q)^{\frac{N_f-N}{2N_f}} v_i z_j^{-1} b^{-1})\,.  
\RRee 
Here ${\bf u}$, ${\bf v}$, and $b$ are parametrizing the $SU(N_f)_u\times  SU(N_f)_v \times U(1)_b$ global symmetry of the theory. on side B we have 
\RRbea
&&{\tilde {\cal I}}_{N_f-N}^{N_f}({\bf u},{\bf v},b)=
\left(\prod_{i,j=1}^{N_f} \Gamma((p q)^{\frac{N_f-N}{N_f}} u_i v_j)\right)\, 
\kappa^{N_f-N-1}\frac1{(N_f-N)!}\oint \prod_{i=1}^{N_f-N-1} \frac{dz_i}{2\pi i z_i}
\prod_{i\neq j} \frac1{\Gamma(z_i/z_j)}\nonumber\\
&& \prod_{i=1}^{N_f}\prod_{j=1}^{N_f-N} \Gamma((p q)^{\frac{N}{2N_f}}b^{\frac{N}{N_f-N}} u_i^{-1} z_j) \Gamma((p q)^{\frac{N}{2N_f}} v_i^{-1} z_j^{-1} b^{-\frac{N}{N_f-N}})\,.  \nonumber\\
\RReea Duality implies that the two indices above should be equal and indeed it was shown by Rains that they are~\cite{RRrains}.  The proof is rather non trivial but in section 7 we will discuss a proof for a certain limit of the parameters.

\subsection{Kutasov-Schwimmer dualities}

Let us give yet another example of duality which implies a mathematical identity of indices yet to be proven rigorously. 
The example is that of Kutasov-Schwimmer dualities where in addition to $N_f$ flavors of fundamental matter of $SU(N)$ gauge group one introduces two, or one, adjoint fields. The superpotentials for the adjoint fields follow $ADE$ classification,
\RRbea
A_k \;:&&\qquad {\rm Tr} \, X^{k+1}\,,\\
D_{k+2}\;:&& \qquad  {\rm Tr} \, X^{k+1} +{\rm Tr} \, XY^2\,,\nonumber\\
E_6\;: &&\qquad {\rm Tr} \, X^4+ {\rm Tr} \,Y^3\,,\qquad E_7\;:\qquad {\rm Tr}\,  X^3 Y+{\rm Tr} \,Y^3\,,\qquad E_8\;:\qquad {\rm Tr }\, X^5+{\rm Tr} \,Y^3\,.\nonumber
\RReea These superpotentials fix the R-charge assignments for the adjoint fields.  Dual descriptions are known in the $A$, $D$,\cite{RRKutasov:1995ve,RRKutasov:1995np,RRIntriligator:2003mi} and $E_7$~\cite{RRKutasov:2014yqa} cases. The dual has gauge group of $SU(\alpha N_f - N)$  with $\alpha$ depending on the superpotential for
the adjoint matter,
\RRbea
A_k \;:&&\qquad \alpha = k\,,\\
D_{k+2}\;:&& \qquad  \alpha = 3k\,,\nonumber\\
E_7\;:&&\qquad \alpha=30\,.\nonumber
\RReea One also has to introduce a variety of singlet fields coupled through a superpotential to gauge singlet operators on the dual side. For details the reader is referred to~\cite{RRKutasov:2014wwa}. One can write down the corresponding identities for 
the supersymmetric indices, see {\it e.g.}~\cite{RRDolan:2008qi}, and check that they are true in series expansion in fugacities or in certain limits such as large $N$. However no proof is known to date.

\

\noindent We have focussed on the simplest representative examples of dualities and  there are many more, see
the discussion in~\cite{RRSpiridonov:2009za,RRSpiridonov:2011hf}. The mathematics of these identities is a very active area of research, see {\it e.g.}~\cite{RRSpiridonov:2010yc,RRSpiridonov2,RRSpiridonov3,RRSpiridonov4} for reviews.

\

\section{Limits}

 In previous sections we have discussed how the index encodes information about  four-dimensional physics. Upon taking appropriate limits,  the index can also be  related to physical quantities in other spacetime dimensions. We will discuss here the two most natural limits of this kind.

\

\subsection{Small \texorpdfstring{$\tau$}{tau} limit,
  \texorpdfstring{$\BS^1\to 0$}{S1 to 0}}

We consider taking all the fugacities to $1$. This limit in the partition function language corresponds to taking the limit 
of the size of $\BS^1$ to zero. Since the index and the partition function differ only by the $e^{-\tau E_{Casimir}}$ factor  the two coincide in the limit. Moreover it was argued on general grounds that in this limit the index has generically the following divergent behavior~\cite{RRClosset:2012ru} 
\RRbe\label{RRanom}
{\cal I}(\tau\to 0)={\cal Z}_{\BS^3\times\BS^1_{\tau}}(\tau\to 0) = e^{-\frac{16\pi^2}3 \frac1{\tau} (a-c)}\times {\cal Z}_{\BS^3}\,. 
\RRee This asymptotic behavior    can be corrected by subleading  power-law terms in $\tau$ when the theory has moduli spaces on the circle \cite{RRArdehali:2015bla,RRAharony:2013kma}.\footnote{However,  in certain non-generic situations even the leading behavior is modified, see~\cite{RRArdehali:2015bla} for a careful discussion. Perhaps the simplest example that exhibits  this non-generic behavior is the ISS model~\cite{RRIntriligator:1994rx} (see also~\cite{RRVartanov:2010xj} for a discussion of the index of this theory).}  
Let us discuss how this comes about in detail in a particular example.\footnote{We follow here the discussion in appendix B of~\cite{RRAharony:2013dha}.}

\

\noindent $\bullet$ {\it Dimensional reduction of the index of a chiral field} --  Let us make the relation between the geometry and the index a bit more precise. We compute the partition function on  $\BS^3\times \BS^1$ with radii $r_3$ and $r_1$, twisted by fugacities for various global symmetries.  Equivalently, after a change of variables it can be thought of as a partition function on $\BS_b^3\times{\widetilde \BS}^1$ with the fugacities responsible for the geometric twisting absorbed in the geometry~\cite{RRImamura:2011wg}. Here is $\BS^3_b$ is the squashed sphere.
We can compute the index as a partition function by first reducing the theory on ${\widetilde \BS}^1$ of finite radius, and then computing the $3d$ partition function of the resulting $3d$ theory, including all the KK modes on the ${\widetilde \BS}^1$.  The fugacities corresponding to flavor symmetries can be thought of as  couplings to background gauge fields along the $\BS^1$ direction. The gauge fields along the $\BS^1$ have the meaning of real mass parameters for global symmetries in three dimensions. In addition, as we go once around the $\BS^1$, we should rotate the $\BS^3$ along the Hopf fiber by an angle depending on the fugacities $p$ and $q$.  This has the effect of changing the geometry.  As discussed in \cite{RRImamura:2011wg}, there is a change of coordinates, where the metric becomes that of an $\BS^3_b \times {\widetilde \BS}^1$, where the ${\widetilde \BS}^1$ factor is rotated on the $\BS^3_b$ base.  The parameters are related by
\RRbe\label{RRpqrel}p=e^{-2\pi\,b \frac{\widetilde r_1}{r_3}} \qquad ; \qquad q=e^{-2\pi\,b^{-1} \frac{\widetilde r_1}{r_3}}\,,\qquad
\tilde r_1=\frac{2}{b+b^{-1}}\,r_1~.\RRee
This procedure leads to the action used in \cite{RRImamura:2011wg} to compute the supersymmetric partition function on $\BS^3_b$. Then, we can write the $4d$ index as coming from a theory on $\BS^3_b$, with an infinite tower of KK modes.
We refer the reader to the references above and to appendix B of \cite{RRAharony:2013dha} for more details.

For a free chiral field (of R-charge $R$ and charged under a $U(1)_u$ symmetry) we are interested in rewriting the index in the following form,
\RRbe\label{RRchiralrefn}
{\cal Z}^{(R)}_{\BS^3\times\BS^1}(p,\,q,\,u)\propto \prod_{n=-\infty}^\infty {\cal Z}^{(R)}(\omega_1,\,\omega_2,\,m+\frac{n}
{{\tilde r}_1})\,,
\RRee where ${\cal Z}_{\BS^3_b}$ is the $\BS^3_b$ partition function of a chiral field depending on the squashing parameter, real mass for $U(1)_u$, and the R-charge,
\RRbea\label{RRhypG}
&&{\cal Z}_{\BS^3_b} =\Gamma_h(\omega\,R+\sum_a m_a\,e_a;\omega_1,\,\omega_2)\,,\\
&&\Gamma_h(z;\omega_1,\,\omega_2)=
e^{\frac{\pi i}{2\omega_1\omega_2}\left((z-\omega)^2-\frac{\omega_1^2+\omega_2^2}{12}\right)}\,
\prod_{\ell=0}^\infty
\frac{1-e^{\frac{2\pi i}{\omega_1}(\omega_2-z)}\,e^{\frac{2\pi i\omega_2\,\ell}{\omega_1}}}
{1-e^{-\frac{2\pi i}{\omega_2}\,z}\,e^{-\frac{2\pi i\omega_1\,\ell}{\omega_2}}}\,.\nonumber
\RReea
The parameters on the two sides in \eqref{RRchiralrefn} are related as
\RRbe
u=e^{2 \pi i{\tilde r}_1\,m},\qquad p=e^{2 \pi i{\tilde r}_1\,\omega_1},\qquad q=e^{2 \pi i{\tilde r}_1 \,\omega_2}\,,\qquad
\omega = \frac12(\omega_1+\omega_2)\,.
\RRee
On the left-hand side we have the $4d$ index of a chiral superfield, and
on the right-hand side the product over $3d$ $\BS^3_b$ partition functions of the KK modes on ${\widetilde \BS}^1$.
The inverse radius of ${\widetilde \BS}^1$, $1/{{\tilde r}_1}$, plays the role of a real mass coupled to the KK momentum.

The expression on the right hand side of \eqref{RRchiralrefn} as it stands is divergent and needs to be properly reguralized and defined. Moreover one needs to be careful to include the Casimir energy in the definition of the partition function in four dimensions.  Concretely, the twisted partition function of the chiral field on $\BS^3\times \BS^1$ can be written as
\RRbe\label{RRchirfieldphase}
{\cal Z}_{\BS^3\times \BS^1}^{(0)}(p,\,q,\,u)=e^{{\cal I}_0}\;\Gamma( u\,;p,\,q)\,.
\RRee Here we chose to take R charge to be zero for simplicity with non trivial R charge easily reintroduced by mixing in the flavor symmetry.
The factor $e^{{\cal I}_0}$ relates the two different natural
 normalizations. It is computed in~\cite{RRKim:2012ava},
\RRbe\label{RRcasimir}
{{\tilde r}_1}^{-1}{\cal I}_0=\frac14
\left.\left({r}^{-1}\frac{d}{dr}\left(r\,\Gamma_0(
e^{2\pi r  i m};e^{2\pi r i \omega_1},e^{2\pi r i\omega_2})\right)\right)\right|_{r=0}\, ,
\RRee
where $\Gamma_0(z;p,q)$ is the so called single particle index, defined by
\RRbe\label{RRpleth}
\Gamma( u\,;p,\,q)=
\exp\left[\sum_{n=1}^\infty\frac1n \; \Gamma_0(z^n;p^n,q^n)\right]\quad\to\quad
\Gamma_0(z;p,q)=\frac{z-pqz^{-1}}{(1-p)(1-q)}\,.
\RRee
Using the fact that $\Gamma_0$ has a simple pole at $r=0$ and a vanishing constant term in the expansion around $r=0$, equation
\eqref{RRcasimir} leads to
\RRbe\label{RRcasimirtwo}{\cal I}_0=
\frac{\pi\,i\,{\tilde r}_1\, (m-\omega) \left(2\, m\,(m - 2 \omega)+ \omega _1\,\omega _2\,\right)}
{6\,\omega _1\, \omega _2}
\,.
\RRee
Next we compute the right-hand side of \eqref{RRchiralrefn}, 
\RRbea\label{RRrhs}
&&\prod_{n=-\infty}^\infty {\cal Z}^{(0)}(\omega_1,\,\omega_2,\,m+\frac{n}{{\tilde r}_1})=
\prod_{n=-\infty}^\infty \Gamma_h(m+\frac{n}{{\tilde r}_1};\omega_1,\omega_2)\,.
\RReea
The infinite product over $n$ here  diverges, since for large $n$ the hyperbolic Gamma functions
approach a divergent exponential behavior, 
\RRbea
&&\log\;\left(  \Gamma_h \left( \omega R + \rho(\sigma) +\tau(\mu+s\,\mu_o) \right) \right) =\\
&&\qquad
{\rm sign}(\tau(\mu_o))\, \frac{\pi i}{2 \omega_1\omega_2} \bigg( \left[\omega (R-1) + \rho(\sigma)+\tau(\mu+s\,\mu_o) \right]^2 - \frac{{\omega_1}^2 + {\omega_2}^2}{12} \bigg) + {\cal O} (e^{-\alpha s}) \,,\nonumber
\RReea
We can regularize this divergence using zeta-function
regularization ($\sum_{n=1}^\infty n^s=\zeta(-s)$)\footnote{
Here we defined ${\rm sign}(n=0)=-1$.
}
\RRbea\label{RRzetaregprod}
\prod_{n=-\infty}^\infty\,
&&e^{-{\rm sign}(n)\,\frac{\pi i}{2\omega_1\omega_2}\left((m+\frac{n}{{{\tilde r}_1}}-\omega)^2-\frac{\omega_1^2+\omega_2^2}{12}\right)}
\longrightarrow  \\
&& \exp\left(\Delta\right)\equiv
\exp\left(
\frac{i \pi \,  \left(2 m (3 m\,{{\tilde r}_1}+1)-2\,(1-6m\,{{\tilde r}_1})\,\omega+{{\tilde r}_1}\,(\omega _1^2+\omega _2^2+3\omega_1\omega_2)\right)}{12\,{{\tilde r}_1}\, \omega _1 \omega _2}
\right)\,.\nonumber
\RReea The precise statement of \eqref{RRchiralrefn} is then the following equality
\RRbe\label{RRregrhs}
e^{{\cal I}_0}\;\Gamma( u\,;p,\,q)=
e^{-\Delta} \;\prod_{n=-\infty}^\infty\,
\left[e^{-{\rm sign}(n)\,\frac{\pi i}{2\omega_1\omega_2}\left((m+\frac{n}{{\tilde r}_1}-\omega)^2-\frac{\omega_1^2+\omega_2^2}{12}\right)}
\Gamma_h(m+\frac{n}{{{\tilde r}_1}};\omega_1,\omega_2)\right]\,.
\RRee The infinite product on the right-hand side is now well-defined, and in fact by using \eqref{RRhypG} and
 \eqref{RRchiralelliptic}
it can be written as a product of two elliptic Gamma functions,
\RRbe\label{RRslthree}
\Gamma(u;p,\,q)=e^{-\Delta-{\cal I}_0}\;
\frac{\Gamma(e^{2\pi i\frac{m}{\omega_1}};
e^{2\pi i\frac{\omega_2}{\omega_1}},e^{-2\pi i\frac{1}{{{\tilde r}_1}\,\omega_1}})}{\Gamma(e^{2\pi i\frac{m-\omega_1}{\omega_2}};e^{-2\pi i\frac{1}{{{\tilde r}_1}\,\omega_2}},e^{-2\pi i\frac{\omega_1}{\omega_2}})}\,.
\RRee
This equality is discussed in \cite{RRfelvar}. It is sometimes viewed as an indication of an $SL(3,\mathbb{Z})$ structure.
Taking the $3d$ limit by sending ${\tilde r}_1$ to zero, we decouple the massive KK modes on the ${\widetilde \BS}^1$.  The only term surviving the limit on the right-hand side of \eqref{RRregrhs} has $n=0$,
and we obtain
\RRbea
\lim_{{\tilde r}_1\to 0}\;
&&\left[\Gamma(e^{2 \pi i {\tilde r}_1\,(\omega\,R+m)};e^{ 2 \pi i {\tilde r}_1 \,\omega_1},\,e^{ 2 \pi i {\tilde r}_1\,\omega_2})\,e^{\frac{\pi i}{6\,\omega_1\,\omega_2\;{\tilde r}_1}\left(m-\omega(1-R)\right)}\right]=\\
&&\qquad\qquad\qquad\qquad\qquad\qquad\qquad
\Gamma_h(\omega\,R+m;\omega_1,\,\omega_2)\,.\nonumber
\RReea Note that the divergent factor is after turning on flavor fugacity and going to unsquashed sphere, $\omega_1=\omega_2=\frac1{2\pi}i$ and $\widetilde r_1=\tau$,
\RRbe
e^{\frac{\pi i}{6\,\omega_1\,\omega_2\;{\tilde r}_1}\omega(1-R)}=
e^{\frac{\pi^2}{3\tau}(1-R)}=e^{-\frac{16\pi^2}{3\tau}(a-c)}\,,
\RRee in agreement with~\eqref{RRanom} since the anomalies of the chiral field are given by
\RRbe 
a= \frac1{32}(9(R-1)^3-3(R-1))\,,\qquad c=\frac1{32}(9(R-1)^3-5(R-1))\,,\qquad\; a-c=\frac1{16} (R-1)\,.
\RRee

\

\noindent $\bullet$ {\it Reduction of gauge theories} -- We  can also consider the limit of small $\tau$ for gauge theories. We will not review this in detail here, and only mention the salient features. Up to the divergent factor appearing in \eqref{RRanom}, and its generalization when flavor fugacities are present, the matrix model for the index reduces to the matrix model~\cite{RRKapustin:2009kz} used to compute $\BS^3_b$ partition function of the dimensionally reduced theory~\cite{RRDolan:2011rp,RRImamura:2011uw,RRGadde:2011ia,RRAharony:2013dha}. Two comments are in order. First, the theories in four dimensions might have classical symmetries which are anomalous in the quantum theory. When reducing the theory on a circle a superpotential is produced which explicitly breaks these symmetries~\cite{RRAharony:2013dha}. In the partition this manifests itself as a lack of real mass parameter for the symmetry which is anomalous in four dimension. These superpotentials are extremely important to understand what physics IR dualities in four dimensions reduce to in three dimensions. Second, in certain cases~\cite{RRAharony:2013kma,RRArdehali:2015bla} the three dimensional partition function in \eqref{RRanom} is by itself divergent. Such examples include reductions of $SO(N)$ gauge theories with ${\cal N}=1$ supersymmetry and $SU(N)$ gauge theories with ${\cal N}=4$ supersymmetry. We refer the reader to \volcite{WI} for details of the $\BS^3_b$ partition function.

\subsection{Large \texorpdfstring{$\tau$}{tau} limit,
  \texorpdfstring{$\BS^3\to 0$}{S3 to 0}}

 In this limit the radius of $\BS^3$ is much smaller than the radius of the circle and we effectively compactify the theory to quantum mechanics on a circle. The supersymmetric index in this limit computes the usual Witten index of the resulting quantum mechanics, that is the number of supersymmetric vacua. More concretely,

\RRbe
{\cal Z}_{\BS^3\times \BS^1_{\tau\to \infty}} \quad\quad  \to \qquad\quad\;\, e^{-\tau E_{\rm Casimir}}\,\#_{\rm vacua}\,.
\RRee In particular since in the index we strip off the Casimir energy contribution it computes in the limit just the number of supersymmetric vacua. However, often a given theory might have a moduli space of vacua and the limit will diverge. In some examples we can keep some of the flavor fugacities which will regulate this divergence and give a finite result.  

For $\tau$  large,\footnote{This limit has also been considered in~\cite{RRSpiridonov:2009za}.} 
\RRbe
p,\;q\;\to\;0\,.
\RRee
We assume implicitly that the index we obtain is finite in the limit because we have enough flavor fugacities to lift the degeneracy of the moduli space (this is not always possible). The fugacities $p$ and $q$  couple to charges $j_2\pm j_1+\frac12 r$. Setting these fugacities to zero is well defined if for all states conrtibuting to the index $j_2\pm j_1+\frac12 r\geq 0$.  Let us assume that this is  the case and soon we will discuss several examples. Then, the states which contribute to the index satisfy
\RRbe
j_2\pm j_1+\frac12 r=0\,,\qquad\to\qquad j_1=0\,\;\qquad  j_2=-\frac12 r\,.
\RRee Moreover, since states contributing to the index satisfy $E-2j_2-\frac32 r=0$ we also get that $E=\frac12 r$. Now, from unitarity,
\RRbe
E\pm 2j_1+\frac32 r\geq 0\,,\qquad\; E\pm 2j_2-\frac32 r\geq0\,,
\RRee which imply that the states contributing to the limit we discuss have all charges vanishing,
\RRbe\label{RRvac}
E=r=j_1=j_2=0\,.
\RRee Such states parametrize vacua of the model, {\it i.e.} the moduli space, as expected. Again, for the index to be well defined we will have to keep some of the flavor fugacities under which the operators contributing to the limit are charged.

Let us discuss the limit for a free chiral field. The limit is well defined if the R-charge is between zero and two. For R-charge vanishing the index is $\frac1{1-u}$ where $u$ is fugacity for the $U(1)$ symmetry rotating the chiral. The index is give just by powers of the scalar component. This is the case when we can give a vacuum expectation value to the scalar parametrizing the moduli space, which will also break the  $U(1)_u$ symmetry. For $r>0$ but less than two the index is $1$. For $r=2$ it becomes $1-u^{-1}$. Note that R-charge two is outside the unitarity bounds for free chiral and thus there is no physical meaning for this result. However, such an R-charge would be acceptable for a gauge non-invariant chiral matter field in gauge theory. 

We now give a more interesting example. Pure $SU(N)$ SYM has $N$ vacua, however it also has a discrete R symmetry.
To define the index we need continuous R symmetry and thus we will not discuss this example but rather turn on flavors.
 Consider $SU(N)$ SQCD with $N_f$ flavors. The standard choice of anomaly R-charge is $\frac{N_f-N}{N_f}$ for all the matter fields. This choice keeps all the flavor symmetry manifest. Our limit is well defined here.
The limit of $p,\, q\to 0$ in this case is trivial, the index is $1$ meaning that only the vacuum in the origin of field space satisfies \eqref{RRvac}. However, we can change the choice of R-charges keeping the condition for R-charges to be anomaly free,
\RRbe
\sum R_i+\sum \widetilde R_i = 2N_f-2N\,.
\RRee For example, let us assign $N$ quarks and $N$ anti-quarks R-charge zero, and the remaining matter R-charge one. The anomaly free condition above is satisfied.  Taking our limit the index becomes
\RRbe
{\cal I}^{(N)}(\{t, \widetilde t\}_{i=1}^N)=\frac1{N!}\oint \prod_{i=1}^{N-1}\frac{dz_i}{2\pi i z_i} \prod_{i\neq j}(1-z_i/z_j) \prod_{i=1}^N \frac1{1-t_i z_i}\frac1{1-{\widetilde t}_i z_i^{-1}}\,.
\RRee Note that $N_f$ does not appear here anymore and there is no condition on fugacities $t_i,\; \widetilde t_i$. This is integral can be easily computed to give
\RRbe\label{RRres}
{\cal I}^{(N)}(\{t,\widetilde t\}_{i=1}^N)=(1-\prod_{i=1}^N t_i\widetilde t_i)\frac1{1-\prod_{i=1}^Nt_i}\frac1{1-\prod_{i=1}^N \widetilde t_i}\prod_{i,j=1}^N\frac1{1-t_i\widetilde t_j}\,.
\RRee This can be easily understood. The product is the product over the mesonic operators surviving the limit parametrizing a slice of the moduli space. The second and third terms are the baryon and the anti-baryon. The first term is an obvious constraint on this moduli space. We see that the index captures neatly a slice of the moduli space of the theory. This is equivalent to the so called Hilbert series of this slice (for discussion of Hilbert series see for example~\cite{RRGray:2008yu,RRHanany:2008kn}).   We can ask how this limit behaves under Seiberg duality. On side B of the duality we will have $SU(N_f-N)$ theory with $N_f$ quarks/anti-quarks and gauge singlets dual to the mesons.
The dual quarks in this case have again R-charges zero and one in our case, now $N$ have R-charge $1$ and $N_f-N$ R-charge zero. The mesons which survive the limit have R-charge zero and R-charges two. Note that as we said above the latter 
cannot be physical because of the violation of unitarity bounds.  The index of the dual theory is
\RRbe
\left[\prod_{i,j=1}^N\frac1{1-t_i\widetilde t_j} \prod_{i,j=N+1}^{N_f}(1-t_i^{-1}\widetilde t_j^{-1})\right] \times {\cal I}^{(N_f-N)}(\{\frac{(\prod_{k=1}^{N_f}t_k)^{\frac1{N_f-N}}}{t_i},\frac{(\prod_{k=1}^{N_f}\widetilde t_k)^{\frac1{N_f-N}}}{{\widetilde t}_i}\}_{i=N+1}^{N_f})\,.
\RRee We can now plug in the result from \eqref{RRres} for ${\cal I}^{(N_f-N)}$  and
\RRbe
\prod_{k=1}^{N_f}\widetilde t_k \prod_{k=1}^{N_f}t_k =1\,,
\RRee from anomaly cancelation to
obtain that the above is equal to
\RRbea
&&\left[\prod_{i,j=1}^N\frac1{1-t_i\widetilde t_j} \prod_{i,j=N+1}^{N_f}(1-t_i^{-1}\widetilde t_j^{-1})\right] \times 
(1-\prod_{i=N+1}^{N_f} t_i^{-1}{\widetilde t_i}^{-1})\\
&&\qquad\qquad \frac1{1-\prod_{k=1}^{N_f}t_k \prod_{i=N+1}^{N_f}t^{-1}_i}\frac1{1-\prod_{k=1}^{N_f}\widetilde t_k\prod_{i=N+1}^{N_f} {\widetilde t_i}^{-1}}\prod_{i,j=N+1}^{N_f}\frac1{1-t_i^{-1}{\widetilde t}_j^{-1}}=\nonumber\\
&& \frac1{1-\prod_{k=1}^{N}t_k }\frac1{1-\prod_{k=1}^{N}\widetilde t_k}
(1-\prod_{i=1}^{N} t_i{\widetilde t_i})\prod_{i,j=1}^N\frac1{1-t_i\widetilde t_j} \,,\nonumber
\RReea in agreement with \eqref{RRres}.

Note that naively it is important in the gauge theory for the limit to be well defined to have the R-charges of all the chiral fields to be between zero and two. However, even if some of the charges of chirals are outside of this region  the limit might be well behaved. Consider for example giving R-charge zero to $N_f+N$ chiral fields and R-charge two to $N_f-N$. This is an anomaly free R-charge. Assuming that $N_f+N$ is even, we might  split the choice above equally between the quarks and anti-quarks, that is giving R-charge zero to $\frac{N_f+N}2$ flavors. In such a case the R-charges of the dual theory are one for $\frac{N_f+N}2$ flavors and $-1$ for $\frac{N_f-N}2$. Thus although naively the limit of the matter is singular from the duality we know that the limit for the gauge invariant operators has to be well defined.

In summary, the $p, q$ to zero limit captures  protected information associated  to a certain submanifold of the moduli space of the theory. The precise submanifold is determined by the choice of the R-charges. One can in principle consider other limits on  fugacities  coupling to combinations of charges which for a given model are non-negative for states contributing to the index. However since the index gets contributions from fermions and bosons in conjugate representations the index would usually get contributions from both negatively and positively charged objects unless the limit is for an R-symmetry. In certain cases the information captured in this limit is equivalent to the Hilbert series of the moduli space. An example is given by~\cite{RRGadde:2011uv} the limit of the index of ${\cal N}=2$ theories corresponding to genus zero Riemann surfaces in class ${\cal S}$ 
terminology~\cite{RRGaiotto:2009we}. See also~\cite{RRRazamat:2014pta,RRCremonesi:2015dja,RRHanany:2015via} for the $3d$ variants of such limits.

\

\subsection{Poles and residues}

The index is a meromorphic function of the fugacities and in general has numerous poles. Let us assume the index has a 
behavior of the following form,
\RRbe
{\cal I}_{0}(a_1,a_2,\cdots) =\frac{{\cal I}_{1}(a_1,a_2,\cdots)}{1-a_1}\,,
\RRee where $a_i$ are some fugaicties and we assume ${\cal I}'$ has no zeros or poles at $a_1=1$. From the trace interpretation of the index  we deduce that there is a bosonic operator in the theory, ${\cal O}$, with charges such that it contributes with weight $a_1$ to the index. Moreover, any power of this operator also contributes to the index. The pole 
at $a_1=1$ corresponds to computing the index while giving weight $1$ to ${\cal O}$. Putting it differently, we consider turning on only fugacities  for symmetries consistent with giving a vacuum expectation value for ${\cal O}$. It is thus natural to interpret the residue of the pole as the index of the theory obtained as the IR fixed point of an RG flow triggered by vacuum expectation value for ${\cal O}$. 

We  have encountered an example of the effect of vacuum expectation values while discussing Higgsing in section \ref{RRgaugeind}. Let us give several additional examples. First let us consider a sigma model with two chiral fields and a superpotential
\RRbe
W=\Phi_1\Phi_2^2\,.
\RRee We have one $U(1)_a$ global symmetry preserved by the superpotential and we choose $\Phi_1$ to have charge $-2$ and $\Phi_2$ has charge $+1$. We also assign R-charge $2R$ to $\Phi_1$ and $1-R$ to $\Phi_2$.  The index of the model is given by
\RRbe
{\cal I}(a)=\Gamma((p q)^{R} a^{-2})\Gamma((p q)^{\frac{1-R}2} a)\,.
\RRee  Note that the chiral ring here has the relations
\RRbe
\Phi_1 \Phi_2 \sim 0\,,\qquad\quad \Phi_2^2\sim 0\,.
\RRee  In particular, as we already discussed not all powers of the scalar component of $\Phi_2$ contribute to the index, but any power of the scalar from $\Phi_1$ appears.  This index has many poles one of which is at $a=(p q)^{\frac{R}2}$. The operator which leads to the divergence is the scalar component of $\Phi_1$.  The residue is given by
\RRbe
{\cal I}(a) \sim \frac1{1-(p q)^{R}a^{-2}}  \Gamma((p q)^{\frac12}) {\cal I}_V^{-1} + O(1)\,.
\RRee The index of $\Phi_2$ becomes $\Gamma((p q)^{\frac12})=1$, which is the index of a massive fields since the vacuum expectation value for $\Phi_1$ generates a mass term for $\Phi_2$. The index of field $\Phi_1$ stripping off the divergence is 
$\Gamma(1)'=\frac1{(q;q)(p;p)}={\cal I}_V^{-1}$ which is the index of the Nambu-Goldstone boson corresponding to the broken $U(1)_a$ symmetry. The residue is thus just given by the index of the Nambu-Goldstone boson as expected as the theory is empty in the IR. It is thus natural to write the general relation
\RRbe
{\cal I}_0(a_1\to 1) =  \frac1{1-a_1} {\cal I}_{IR}(a_2,\cdots) {\cal I}_{Nam.-Gold.} + O(1)\,.
\RRee 

\

\begin{figure}[ht]
\begin{center}
\includegraphics[scale=.41]{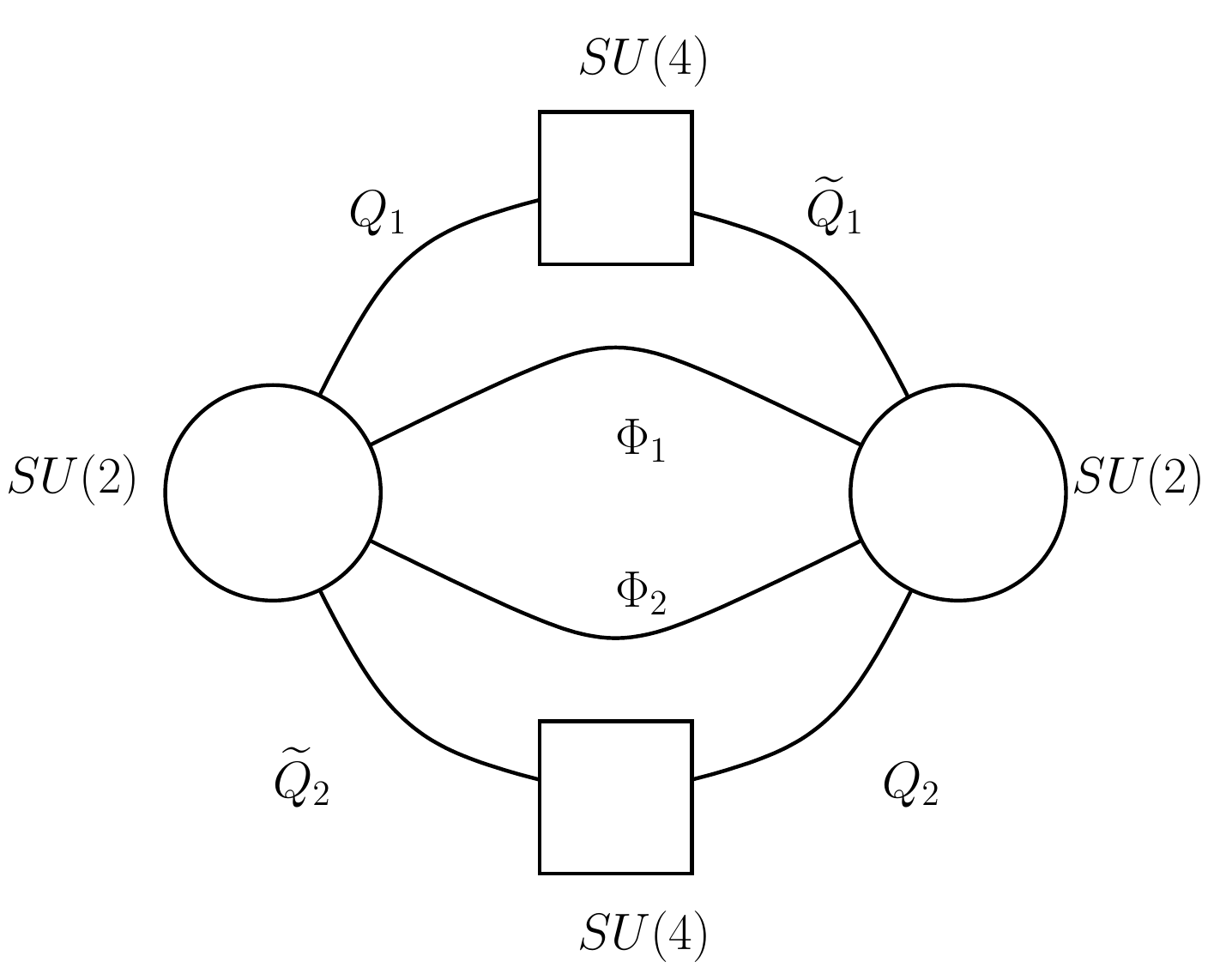}
\end{center}
\caption{An $SU(2)\times SU(2)$ quiver gauge theory.}\label{RRquifi}
\end{figure}

We consider now a more involved  example of a gauge theory. The theory we discuss is $SU(2)\times SU(2)$ quiver gauge theory of figure \ref{RRquifi}.
The superpotential is \RRbe
W=Q_1\Phi_1 \widetilde Q_1+Q_2\Phi_2\widetilde Q_2\,.
\RRee We will assign R-charge zero to the $Q_i$ and $\widetilde Q_i$ fields and R-charge two to $\Phi_j$. This model has three abelian global symmetries which we will denote by $U(1)_T\times U(1)_X\times U(1)_Y$. The different fields have the charges specified in Table 3.
\begin{center}
\begin{table}[h!]
{\small
\begin{centering}
\begin{tabular}{|l|l|l|l|l|}
\hline
 & $U(1)_T$ & $U(1)_X$ & $U(1)_Y$ & $U(1)_R$\tabularnewline
\hline
$\Phi_1$ & $-2$ & $+2$ & $\;0$ & $\;2$\tabularnewline
$\Phi_2$ & $-2$ & $-2$ & $\;0$ & $\;2$\tabularnewline
$Q_1$ & $+1$ & $-1$ & $+1$ & $\;0$\tabularnewline
$\widetilde Q_1$ & $+1$ & $-1$ & $-1$ & $\;0$\tabularnewline
$Q_2$ & $+1$ & $+1$ & $-1$ & $\;0$\tabularnewline
$\widetilde Q_2$ & $+1$ & $+1$ & $+1$ & $\;0$\tabularnewline
\hline
\end{tabular}
\par\end{centering}
} \caption{\label{RRchs}Abelian charges of fields.}
\end{table}
\par\end{center}\vspace{-15pt}
We can consider giving a vacuum expectation value to a baryonic operator of the form $B_Q= \epsilon \cdot Q_1^2$. This will Higgs one of the $SU(2)$ gauge groups and reduce the rank of the flavor group by one. Let us analyze how this comes about from the index. The index of the model is given by,

\RRbea
&&{\cal I}= \kappa^2\oint \frac{dz_1}{4\pi i z_1 \Gamma(z_1^{\pm1})}\oint \frac{dz_2}{4\pi i z_2\Gamma(z_2^{\pm2})}
\Gamma(\frac{p q}{T^2} X^{\pm2} z_1^{\pm1}z_2^{\pm1}) \times\\
&& \;\;\;\;\;\prod_{i=1}^4\Gamma(\frac{T Y}X u_iz_1^{\pm1})\Gamma(T Y X v_i^{-1}z_2^{\pm1})
\Gamma(\frac{T}{Y X} u_i^{-1}z_2^{\pm1}) \Gamma(\frac{T X}Y v_i z_1^{\pm1})\,.\nonumber
\RReea We have two $SU(4)$ symmetries with fugacities $u_i$ and $v_i$.
Baryon $B_Q$ contributes to the index with weight $\frac{T^2 Y^2}{X^2} u_1 u_2$ where we have made a choice of the subgroup of $SU(4)_u$ under which $B_Q$ is charged. Giving a vacuum expectation value to $B_Q$ we set the weight of the operator to one,

$$\frac{T^2 Y^2}{X^2} u_1 u_2=1\,.$$ With this specification of parameters the field $Q_1$ contributes to the index as,

\RRbe
\Gamma(\frac{T Y}X u_iz_1^{\pm1}) \;\;\;\to \;\; \Gamma(\frac1{\sqrt{u_1u_2}}u_iz_1^{\pm1})\,.
\RRee Before the spcification this field had poles in $z_1$ at, among others, $$z_1= \frac{T Y}X u_1, \frac{T Y}X u_2,\qquad\quad\;\;\;
z_1=\frac{X}{T Y} u_1^{-1}, \frac{X}{T Y} u_2^{-1}\,, $$ with the former inside the unit circle and latter outside assuming that $|T|<1$ and all other fugacities for global symmetries being phases.  The integration contour thus lies between these poles.
However after the specification the poles above inside and outside of the circle collide and pinch the integration contour at $z_1=\sqrt{\frac{u_1}{u_2}}, \sqrt{\frac{u_2}{u_1}}$. This causes the integral over $z_1$ to diverge. The residue of this divergence is,

\RRbea
&&2\kappa Res_{\frac{T^2 Y^2}{X^2} u_1 u_2\to1} {\cal I} = 
\prod_{i=3}^4\Gamma(\frac{u_i}{u_1})\Gamma(\frac{u_i}{u_2})\prod_{i=1}^4 \Gamma(T^2 v_iu_1)\Gamma(T^2 v_iu_2)\times\\
&&\;\; \kappa
\oint \frac{dz_2}{4\pi i z_2\Gamma(z_2^{\pm2})}
\Gamma(\frac{p q}{T^4Y^2}\frac1{ u_1 u_2} (\sqrt{\frac{u_1}{u_2}})^{\pm1}z_2^{\pm1}) 
\prod_{i=3}^4\Gamma(\frac{1}{Y^2} \frac{u_i^{-1}}{\sqrt{u_1u_2} }z_2^{\pm1}) \prod_{i=1}^4\Gamma(T^2 Y^2v_i^{-1}\sqrt{u_1u_2}z_2^{\pm1})
\,.\nonumber
\RReea
This is the index of ${\cal N}=1$ $SU(2)$ SCFT with four flavors and additional singlet fields coupled to the charged matter through a superpotential.  This is exactly the matter content one would expect after giving a vacuum expectation value to baryon $B_Q$.  

\

More general poles correspond to turning on vacuum expectation values to derivatives of operators and thus break explicitly Lorentz invariance. The theory in the IR is expected to have co-dimension two defects. The residue computes then an index of a theory in presence of such defects. Such flows and corresponding defects were discussed in the ${\cal N}=2$ context in~\cite{RRGaiotto:2012xa} and in ${\cal N}=1$ context in~\cite{RRGaiotto:2015usa} (see~\cite{RRRastelli:2014jja} for a review). The IR theory here has  $4d$ degrees of freedom coupled to $2d$ ones, and the index is often expressible as some difference operator, shifting flavor fugacities by general powers of $p$ and $q$, acting on the four dimensional index~\cite{RRGaiotto:2012xa,RR2014JHEP...03..080G,RRAlday:2013kda,RRGaiotto:2015usa}. This is reminiscent of the observation below \eqref{RRindtt}.

\

\subsection{Large $N$ limit}

The matrix models of indices of gauge theories can be simplified and explicitly evaluated in the limit of large number of colors using  large $N$  matrix model techniques (see {\it e.g.}~\cite{RRAharony:2003sx,RRKinney:2005ej}). Let us here give a general result for the large $N$ limit of an index of a quiver gauge theory with $U(N)$ gauge groups. We follow here the 
discussion and notations of~\cite{RRGadde:2010en}.

We  consider a quiver theory with gauge group $\prod_{a=1}^sU(N_a)_{u_a}$. Let
$\{e^{\alpha_{ai}}\}_{i=1}^{N_{a}}$ denote the $N_{a}$ eigenvalues
of $u_{a}$. Then the matrix model integral \eqref{RRintegral} is,
\begin{equation}
{\cal I}(x)=\int\prod_{a,i}[d\alpha_{ai}]\exp\left\{ -\sum_{ai\neq bj}V_{b}^{a}(\alpha_{ai}-\alpha_{bj})\right\}\,.
\end{equation}
Here, the potential $V$ is the following function
\begin{equation}
V_{b}^{a}(\theta)=\delta_{b}^{a}(\ln2)+\sum_{n=1}^{\infty}\frac{1}{n}[\delta_{b}^{a}-i_{b}^{a}(x^{n})]\cos n\theta\,,
\end{equation}
where, $i_{b}^{a}(x)$ is the total single letter index in the representation
$r^{a}\otimes r_{b}$ and $x$ stands for all the fugacities we can turn on. Writing the density of the eigenvalues $\{e^{\alpha_{ai}}\}$
at the point $\theta$ on the circle as $\rho_{a}(\theta)$, we reduce
it to the functional integral problem,
\begin{equation}
{\cal I}(x)=\int\prod_{a}[d\rho_{a}]\exp\{-\int d\theta_{1}d\theta_{2}\sum_{a,b}n_{a}n_{b}\rho_{a}(\theta_{1})V_{b}^{a}(\theta_{1}-\theta_{2})\rho_{b}^{\dagger}(\theta_{2})\}\,.
\end{equation}
For large $N$, we can evaluate this expression with the saddle point
approximation,\begin{eqnarray*}
{\cal I}(x) & = &  \prod_{k}\frac{1}{\det(1-i(x^{k}))}\,.\end{eqnarray*}
For $SU(N)$ gauge groups instead of $U(N)$, the result is modified
as follows,
\begin{equation}
{\cal I}(x)=\prod_{k}\frac{e^{-\frac{1}{k}\mbox{tr }i(x^{k})}}{\det(1-i(x^{k}))}\,.
\end{equation}
 Here  $i(x)$ is the  matrix with entries $i_{b}^{a}(x)$.

The single-trace partition function can be obtained from the
full partition function,
\begin{eqnarray}
{\cal I}_{s.t.} & = & \sum_{n=1}^{\infty}\frac{\mu(n)}{n}\log{\cal I}(x^{n})\\
 & = & -\sum_{k=1}^{\infty}\frac{\varphi(k)}{k}\log[\det(1-i(x^{k}))]-\sum_{n=1}^{\infty}\frac{\mu(n)}{n}\sum_{k=1}^{\infty} \frac{\mbox{tr }i(x^{n k})}{k}\\
 & = & -\sum_{k=1}^{\infty}\frac{\varphi(k)}{k}\log[\det(1-i(x^{k}))]-\mbox{tr }i(x)\,.
 \end{eqnarray}
The second term in the summation would be absent for the $U(N)$ gauge theories.
Here $\mu(n)$ is the M\"obius function ($\mu(1)\equiv1,\:\mu(n)\equiv0$
if $n$ has repeated prime factors and $\mu(n)\equiv(-1)^{k}$ if
$n$ is the product of $k$ distinct primes) and $\varphi(n)$ is the
Euler Phi function, defined as the number of positive integers less
than $n$ that are coprime to $n$. We have used the properties
\begin{equation}
\sum_{d|n}d\mu(\frac{n}{d})=\varphi (n),\qquad\qquad\sum_{d | n}\mu(d)=\delta_{n,1}.
\end{equation}

Indices in the large $N$ limit can be used to check holographic dualities. For example the index of ${\cal N}=4$ SYM in this limit can be matched with the spectrum of fields in $AdS_5$ computed in supergravity~\cite{RRKinney:2005ej}. The large $N$ indices~\cite{RRGadde:2010en} of a variety of $Y_{p,q}$ models~\cite{RRBenvenuti:2004dy} where also matched with the holographic duals~\cite{RREager:2012hx}. In general the field theory expressions in the large $N$ limit are rather simple though the dual holographic computation can be involved, see~\cite{RREager:2012hx}. For example, the index of ${\cal N}=2$ class ${\cal S}$ theories~\cite{RRGaiotto:2009we} of genus ${\frak g}$ is explicitly known in large $N$ limit~\cite{RRGadde:2011uv} though that simple result was not yet reproduced from the gravity side~\cite{RRGaiotto:2009gz}.

\

\section{Other topics and open problems}

There are many other interesting related topics that we could review here. We conclude with a brief mention of a few of them:
\begin{itemize}
\item {\it Holomorphic blocks} --  The localization procedure leading directly to the trace-formula formulation of the index is the so called Coulomb branch localization. The computation reduces to a matrix integral over the zero modes of the vector 
field in the direction of $\BS^1_\tau$. The name comes from the fact that these components upon reduction to three dimensions become scalar components in the vector multiplet and parametrize the Coulomb branch. However, there is a different localization procedure one can employ~\cite{RRPeelaers:2014ima,RRYoshida:2014qwa,RRNieri:2015yia}. The dimensional reduction of this procedure to three dimensions leads to  the so called Higgs branch localization form for the index~\cite{RRPasquetti:2011fj,RRBeem:2012mb,RRBenini:2013yva}. In this localization procedure the index can be written as a finite sum over vortex/anti-vortex partition functions which are effectively partition functions on ${\mathbb C}\times  T^2$. This ``holomorphic block'' factorization of the partition function is extremely powerful since it connects together apriori unrelated partition functions. By gluing differently the blocks one can obtain various geometry and thus relate the supersymmetric index for example to $\BS^2\times T^2$ partition function. Let us mention here only the simplest example of such a factorization in the case of a free chiral field. Here we have
\RRbe\label{RRfacc}
{\cal I}^{(R)}(a) =\Gamma((pq)^{\frac{R}2} a;p,q) = 
\Gamma((p q)^{\frac{R}2} a;p,pq)
\Gamma((p q)^{\frac{R}2} q a;q,pq)\,.
\RRee There are many interesting results yet to be uncovered following this direction.

\

\item {\it Lens space index} -- As was mentioned in the introduction the supersymmetric index is a  special case of a sequence of partition functions, the lens space indices $\BS^3/\Zb_r \times \BS^1$~\cite{RRBenini:2011nc}.  As a counting problem 
the lens index is computed as follows. Since the geometry involves an orbifold projection the lens index receives contributions from local operators consistent with the action of the orbifold. Let us call this sector the ``untwisted'' one. 
Let us again here give just an example of the lens index of a free chiral field in the ``untwisted'' sector,

\RRbe
{\cal I}^{(R)}_{r}(a) = \Gamma((p q)^{\frac{R}2} a;p^r,pq)
\Gamma((p q)^{\frac{R}2} q^r a;q^r,pq)\,.
\RRee
On the other hand, for $r>1$ the lens space $\BS^3/\Zb_r$ has a non-contractable torsion cycle, and upon quantizing the theory on this space one should consider configurations wrapping this cycle. This leads to a finite number, since the cycle is torsion, of ``twisted'' sectors which receive contributions from extended objects in the theory. Thus although the supersymmetric index, $r=1$, gets contributions only from local operators, the lens index captures a much larger variety of objects. Moreover, the spectrum of the non-local objects is sensitive to the global structure of the gauge groups~\cite{RRRazamat:2013opa} and not just to the Lie algebras making lens indices a more refined characteristic of the physics. 
Taking the limit of large $r$ the non-trivial cycle of the lens space shrinks to zero size and $\BS^3/\Zb_r$ becomes $\BS^2$. 
In this limit the lens index in four dimensions reduces to the supersymmetric index in three dimensions. The finite sum over the twisted sectors becomes an infinite sum over monopoles sectors in three dimensions. Although there are several works 
studying the lens index it has been largely neglected and there are many avenues for farther research.

\

\item {\it Relations to integrable models} -- Finally let us mention that the supersymmetric index is closely related to quantum mechanical integrable systems.  These relations come in different forms. For example the (lens) index itself can be related to partition function of two dimensional lattice integrable models~\cite{RRSpiridonov:2010em,RRYamazaki:2013nra}. On the other hand, as we discussed in the previous sections, computing indices of theories in presence of surface defects amounts to acting on indices without defects with difference operators~\cite{RRGaiotto:2012xa,RRGaiotto:2014ina,RRGaiotto:2015usa}. Such difference operators are Hamiltonians for well known 
Ruijsenaars-Schneider integrable systems when the theories are ${\cal N}=2$~\cite{RRGaiotto:2012xa,RRAlday:2013kda,RRBullimore:2014nla,RRAlday:2013rs,RRRazamat:2013jxa}, and give rise to novel integrable models when one has ${\cal N}=1$ supersymmetry~\cite{RRGaiotto:2015usa,Yagi:2015lha,Maruyoshi:2016caf}. These relations deserve a much  more thorough investigation.

\end{itemize}

\section*{Acknowledgements}

We would like to thank Ofer Aharony, Chris Beem, Abhijit Gadde, Davide Gaiotto, Guido Festuccia, Zohar Komargodski, Nathan Seiberg, Brian Willett, Wenbin Yan  for fruitful collaborations and numerous discussions on the topics reviewed here. LR is supported in part by   NSF Grant PHY-1316617. SSR is  a Jacques Lewiner Career Advancement Chair fellow. This research was also supported by Israel Science Foundation under grant no. 1696/15 and by I-CORE  Program of the Planning and Budgeting Committee.

\documentfinish